\documentclass[12pt]{iopart}

\pdfoutput=1

\usepackage{graphicx,epsfig,sidecap}
\usepackage{bm}
\usepackage{braket}
\usepackage{amssymb}
\usepackage{amsfonts}
\usepackage{mathrsfs}
\usepackage{amsmath}
\usepackage{graphicx}
\usepackage{bm}
\usepackage{xcolor}
\usepackage{hyperref}
\usepackage{enumerate}
\usepackage[toc,title]{appendix}

\catcode`@=11 
\renewcommand\tableofcontents{%
  \section*{\contentsname}%
  \@starttoc{toc}%
}
\catcode`@=12

\def\be{\begin{equation}}
\def\ee{\end{equation}}

\def\bea{\begin{eqnarray}}
\def\eea{\end{eqnarray}}

\def\Tr{{\rm Tr}}

\numberwithin{equation}{section}

\begin{document}

\title[
Extrapolating entanglement entropy and negativity of disjoint intervals in CFT
]{
\\
Entanglement entropy and negativity of 
disjoint intervals in CFT: Some numerical extrapolations
}

\vspace{.7cm}

\author{Cristiano De Nobili, Andrea Coser and Erik Tonni}
\address{SISSA and INFN, via Bonomea 265, 34136 Trieste, Italy. \\}

\vspace{.8cm}

\begin{abstract}
The entanglement entropy and the logarithmic negativity can be computed in quantum field theory through a method based on the replica limit. 
Performing these analytic continuations in some cases is beyond our current knowledge, even for simple models. 
We employ a numerical method based on rational interpolations to extrapolate the entanglement entropy of two disjoint intervals for the conformal field theories given by the free compact boson and the Ising model. 
The case of three disjoint intervals is studied for the Ising model and the non compact free massless boson. For the latter model, the logarithmic negativity of two disjoint intervals has been also considered. 
Some of our findings have been checked against existing numerical results obtained from the corresponding lattice models. 
\end{abstract}

\maketitle

\newpage


\section{Introduction}
\label{sec intro}

Entanglement measures have been the focus of an intense research activity in condensed matter theory, quantum information, quantum field theory and quantum gravity during the last decade. 
The most celebrated one among them is the entanglement entropy, which measures the entanglement between two complementary parts when the whole system is in a pure state \cite{rev}. 
Considering a quantum system in its ground state $|\Psi \rangle$, or in any other pure state, and assuming that its Hilbert space is factorized as $\mathcal{H}= \mathcal{H}_A \otimes \mathcal{H}_B$, the $A$'s reduced density matrix is defined as $\rho_A \equiv \Tr_B \rho$, being $\rho = |\Psi \rangle \langle \Psi |$ the density matrix of the whole system. 
The reduced density matrix $\rho_A$, which characterizes a mixed state, is normalized by requiring that $\Tr_A \rho_A =1$.
The entanglement entropy $S_A$ is the Von Neumann entropy associated to $\rho_A$.  Analogously, one can introduce $S_B$ and, since $\rho$ corresponds to a pure quantum state, we have that $S_B =  S_A$.
In quantum field theory, the entanglement entropy is usually computed by employing the replica limit, namely
\be
\label{EEdef replica}
S_A 
\equiv -\, \Tr (\rho_A \log \rho_A)
= \lim_{n \to 1} S_A^{(n)} \,,
\ee
where $S_A^{(n)} $ are the R\'enyi entropies, which are defined as follows
\be
\label{renyi def}
S_A^{(n)} \equiv \frac{\log \Tr(\rho_A^n)}{1-n}\,.
\ee
From this expression and the normalization condition for $\rho_A$, it is straightforward to find that $S_A = - \partial_n \Tr(\rho_A^n) |_{n=1}$.
Typically, $S_A^{(n)}$ is known for positive integers $n$ and therefore it must be analytically continued to real values of $n$ in order to perform the replica limit (\ref{EEdef replica}).

In quantum field theory, the entanglement entropy is a divergent quantity when $a \to 0$, being $a$ the UV cutoff. In many cases the coefficient of the leading divergence is proportional to the area of $\partial A$ and this property is known as the area law for the entanglement entropy. 
This rule has some important exceptions and the main one is a generic two dimensional conformal field theory (CFT) at zero temperature. Considering an infinite line and an interval of length $\ell$ as the subsystem $A$, we have that $\partial A$ is made by the two endpoints of the interval and it is well known that $S_A = (c/3) \log(\ell/a) +\textrm{const}$, where $c$ is the central charge of the model \cite{Holzhey, cc-04} (see also \cite{cc-rev} for a review).

\begin{figure}[t]
\vspace{.4cm}
\hspace{.8cm}
\includegraphics[width=15cm]{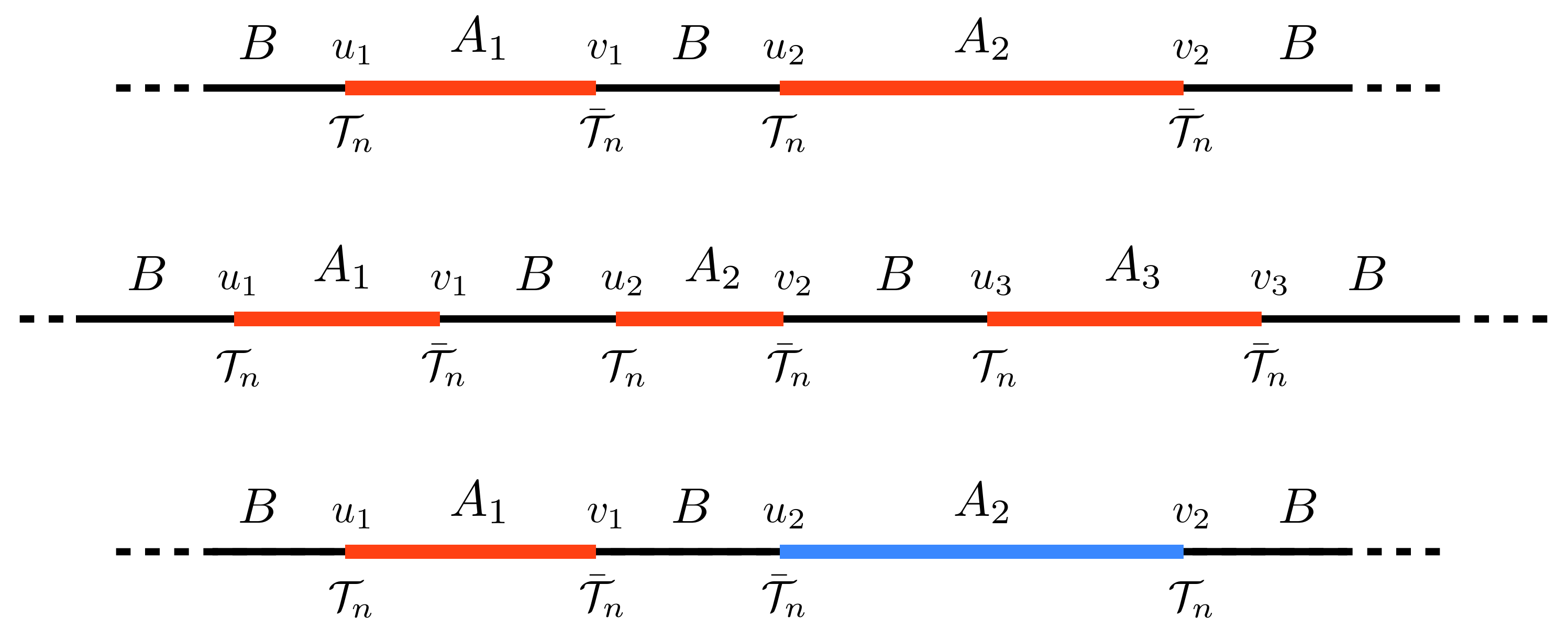}
\vspace{.1cm}
\caption{
The configurations of intervals considered. Top and middle: the entanglement between a subsystem $A$ made by either two (top) or three (middle) disjoint intervals and the remainder $B$. Bottom: the entanglement between two disjoint intervals $A_1$ and $A_2$ embedded in a larger system in its ground state made by $A_1 \cup A_2$ and the reminder $B$.
In CFT correlation functions of branch-point twist fields $ \mathcal{T}_n$ and $ \bar{\mathcal{T}}_n$ placed at the endpoints of the intervals must be computed to get either the entanglement entropy (top and middle panels) or logarithmic negativity (bottom panel) through the proper replica limit.   
}
\label{fig:intro}
\end{figure}

An important configuration to study is when the subsystem $A = A_1 \cup A_2$ is made by two disjoint spatial regions $A_1$ and $A_2$ (see Fig.\,\ref{fig:intro}, top panel, for one spatial dimension).
In this case, it is convenient to introduce the mutual information, which is defined as
\be\label{MIdef}
I_{A_1,A_2}
\equiv
S_{A_1}+S_{A_2}-S_{A_1 \cup A_2}
= 
\lim_{n \to 1} I_{A_1,A_2}^{(n)}\,,
\ee
where in the last step we have emphasized that $I_{A_1,A_2}$ can be found as the replica limit of the following combination of R\'enyi entropies
\be
\label{RenyiMI}
I_{A_1,A_2}^{(n)}
\equiv 
S^{(n)}_{A_1}+S^{(n)}_{A_2}-S^{(n)}_{A_1 \cup A_2}
=
\frac{1}{n-1}\,\log R_{2,n} \,,
\qquad
R_{2,n} \equiv \frac{\Tr \rho^n_{A_1\cup A_2}}{\Tr \rho^n_{A_1} \Tr \rho^n_{A_2}}\,.
\ee
The subadditivity of the entanglement entropy guarantees that $I_{A_1,A_2} \geqslant 0$ and the leading divergence of the different terms cancels in the combination (\ref{MIdef}) when the area law holds. 
Moreover, the mutual information (\ref{MIdef}) could contain more physical information with respect to the entanglement entropy of a single region. For instance, in two dimensional CFTs, while $S_A$ of a single interval depends only on the central charge, the mutual information $I_{A_1,A_2}$ encodes all the CFT data of the model (conformal dimensions of the primaries and OPE coefficients)  \cite{cg-08, fps-08, cct-09, cct-11, atc-10, fc-10, many-int}. The mutual information has been studied also through the holographic approach \cite{RT, headrick}.

Taking the limit $n \to 1$ in (\ref{EEdef replica}) and (\ref{MIdef}) in many interesting cases is highly non trivial.
For instance, the analytic continuation of the R\'enyi entropies of a single interval for the excited states given by the primaries \cite{abs-excited, excited-bdy} has been studied in \cite{cfl-excited}. For the excited states given by the descendants a closed expression for all the R\'enyi entropies is still not known \cite{p-14}.
Interesting features have been observed by considering the R\'enyi entropies of a single interval in critical one dimensional models for real $n$ but no singularities have been found \cite{gt-10}.

In this paper we address the case of disjoint intervals for some models in one spatial dimension. 
The R\'enyi entropies for a subsystem $A$ made by $N$ disjoint intervals (see Fig.\,\ref{fig:intro}, middle panel for $N=3$) are given by the partition function of the model on a Riemann surface of genus $g=(N-1)(n-1)$. 
These partition functions can be computed for some simple CFTs like the massless compact  boson and the Ising model \cite{cct-09, cct-11, ctt-14} but finding the corresponding analytic continuations in the most generic case is still beyond our knowledge. 
For two spatial dimensions, already the simple case of the entanglement entropy of a disk could lead to a difficult replica limit \cite{ahjk-14}.

Another interesting quantity to consider is the logarithmic negativity, which is a measure of entanglement for bipartite mixed states \cite{log-neg-vari}.
Let us consider a pure or mixed state characterized by the density matrix $\rho$ acting on a bipartite Hilbert space $\mathcal{H}= \mathcal{H}_1 \otimes \mathcal{H}_2$ and  the arbitrary bases $| e_i^{(1)} \rangle$ and $| e_j^{(2)} \rangle$ for $\mathcal{H}_1$ and $\mathcal{H}_2$ respectively.
The important object to introduce is the partial transpose of $\rho_{A_1\cup A_2}$ with respect to one of the two parts. Considering e.g. the partial transposition with respect to the second part, the matrix element of $\rho_{A_1\cup A_2}^{T_2}$ is defined as follows
\be
\label{PartialTransp}
\langle e_i^{(1)} e_j^{(2)}| \rho_{A_1\cup A_2}^{T_2}| e_k^{(1)} e_l^{(2)}\rangle = \langle e_i^{(1)} e_l^{(2)}| \rho_{A_1\cup A_2}| e_k^{(1)} e_j^{(2)}\rangle\,.
\ee
Then, the logarithmic negativity is given by
\be
\label{logneg def}
\mathcal{E}
\equiv
\log \Tr |\rho_{A_1\cup A_2}^{T_2}|\,,
\ee
where $\Tr |\rho_{A_1\cup A_2}^{T_2}|$ is the trace norm of the hermitean matrix $\rho_{A_1\cup A_2}^{T_2}$, which is the sum of the absolute values of its eigenvalues. 
Taking into account the traces $\Tr (\rho_{A_1\cup A_2}^{T_2} )^n$ of integer powers of $\rho_{A_1\cup A_2}^{T_2}$, it is not difficult to observe that a parity effect occurs.
In particular, considering the sequence of the odd powers $n=n_o$ and the one of the  
even powers $n=n_e$, the logarithmic negativity (\ref{logneg def}) can be found through the following replica limit \cite{cct-neg-letter,cct-neg-long}
\be
\label{replica limit neg}
\mathcal{E} = \lim_{n_e \to 1} 
\log \Tr \big(\rho_{A_1\cup A_2}^{T_2} \big)^{n_e} .
\ee
Notice that for $n_o \to 1$ one simply recovers the normalization condition $\Tr \rho_{A_1\cup A_2}^{T_2} =1$.
For a bipartite pure state a relation occurs between $\Tr (\rho_{A_1\cup A_2}^{T_2} )^n$ and the Renyi entropies which tells us that the logarithmic negativity reduces to the R\'enyi entropy of order $n=1/2$. 
However, we are interested in the logarithmic negativity of mixed states and the reduced density matrix is an important example.
Thus, given a quantum system in a pure state and considering the reduced density matrix $\rho_{A_1\cup A_2}$ of two adjacent or disjoint spatial regions, while $S_{A_1\cup A_2}$ measures the entanglement between $A_1\cup A_2$ and the complementary region $B$, the logarithmic negativity in (\ref{logneg def}) measures the entanglement between $A_1$ and $A_2$ (see Fig.\,\ref{fig:intro}, bottom panel, for one spatial dimension). 

In two dimensional CFTs, the logarithmic negativity has been studied in \cite{cct-neg-letter, cct-neg-long} for zero temperature, at finite temperature  \cite{cct-neg-T} and also out of equilibrium (the time evolution after a global quench \cite{neg-quench-global}  and after a local quench \cite{neg-quench-local} have been considered).
 For two disjoint intervals at zero temperature $\Tr (\rho_{A_1\cup A_2}^{T_2} )^n$ must be computed case by case because it encodes all the CFT data.
The replica limit (\ref{replica limit neg}) for these expressions turns out to be difficult to compute, like for the mutual information. Indeed, analytic results have not been found for all the possible configurations of intervals.

In this paper we numerically extrapolate the entanglement entropy and the logarithmic negativity through their replica limits, which are respectively (\ref{EEdef replica}) and (\ref{replica limit neg}), for simple two dimensional CFT models and for configurations of intervals whose analytic continuations for $S_A$ and $\mathcal{E}$ are not known. 
In particular, for the free massless boson, both compactified and in the decompactification regime, and for the Ising model, $\Tr \rho_A^n$ are known analytically for a generic number $N$ of disjoint intervals \cite{cct-09, cct-11, ctt-14}, while $\Tr (\rho_{A_1\cup A_2}^{T_2} )^n$ is known analytically for two disjoint intervals \cite{cct-neg-letter, cct-neg-long, a-13, ctt-13}.
We consider some of these models for two or three disjoint intervals (only some configurations in the latter case) and employ a numerical method based on rational interpolations to get the corresponding entanglement entropy or logarithmic negativity. 
This extrapolating method has been first suggested in this context by \cite{ahjk-14} (see \cite{book-numerics} for other numerical methods).
We checked our extrapolations against numerical results found through the corresponding lattice models whenever they are available in the literature, finding very good agreement; otherwise the method provides numerical predictions that could be useful benchmarks for future studies.

The paper is organized as follows. 
In \S\ref{sec MI} we extrapolate the mutual information for the compact boson and for the Ising model comparing the results with the corresponding ones found for the XXZ spin chain \cite{fps-08} and the critical Ising chain \cite{atc-10}.
In \S\ref{sec tripartite} the entanglement entropy of three disjoint intervals is considered for the non compact boson and for the Ising model. 
While the extrapolations for the former model can be checked against exact results for  the periodic harmonic chain, there are no results in the literature about the entanglement entropy of three disjoint intervals for the critical Ising chain to compare with.
In \S\ref{sec neg 2int} we focus on the logarithmic negativity of two disjoint intervals for the non compact boson.
The appendix \S\ref{RIsection} contains a discussion about the rational interpolation method that has been employed throughout the paper.

\section{Mutual information}
\label{sec MI}

In this section, after a quick review of the computation of $I_{A_1,A_2}^{(n)}$ in CFT, we focus on the compactified boson and on the Ising model because $I_{A_1,A_2}^{(n)}$ is known analytically in these cases. 
The numerical extrapolation of the analytic expressions for $I_{A_1,A_2}^{(n)}$ to $n \to 1$ leads to the mutual information, which can be compared with the corresponding  numerical results found from the XXZ spin chain and the Ising chain in a transverse field.

Let us consider a two dimensional CFT with central charge $c$ at zero temperature.\\
As first discussed in \cite{cc-04}, $\Tr \rho_A^n$ for a subsystem $A$ made by $N$ disjoint intervals can be computed as the $2N$-point correlation function of {\it branch-point twist fields} $\mathcal{T}_n$ and $\bar{\mathcal{T}}_n$ placed at the endpoints of the intervals in an alternate sequence (see \cite{ccd-09} for integrable quantum field theories).
These fields have been largely studied in the early days of string theory \cite{bosonization higher genus} and their crucial role for the entanglement computations has been exploited during the last decade.

When the subsystem $A$ is a single interval $A = [u, v]$ with length $\ell = |u-v|$ on the infinite line, $\Tr \rho_A^n$ is given by the two-point function of branch-point twist fields \cite{cc-04}
\be
\label{twist 2point func}
\Tr \rho_A^n  
= 
\langle  \mathcal{T}_n(u)  \bar{\mathcal{T}}_n(v)  \rangle 
=
\frac{c_n}{|u-v|^{2\Delta_n}}\,,
\qquad
\Delta_n = \frac{c}{12} \left( n-\frac{1}{n} \right),
\ee
where $\Delta_n$ are the scaling dimensions of the twist fields $\mathcal{T}_n$ and $\bar{\mathcal{T}}_n$, being $c_n$ a non universal constant such that $c_1=1$.
Taking the replica limit (\ref{EEdef replica}) of (\ref{twist 2point func}) is straightforward and one gets the well known result for the entanglement entropy of an interval in the infinite line \cite{Holzhey}
\be
S_A = \frac{c}{3} \,\log(\ell/a) + c'_1\,,
\ee
where $a$ is a UV cutoff.
Thus, the entanglement entropy and the R\'enyi entropies for a single interval depend only on the central charge $c$ of the model.

When the subsystem $A =A_1 \cup A_2$ is made by two disjoint intervals $A_1 = [u_1, v_1]$ and $A_2 = [u_2, v_2]$ (with the endpoints ordered as $u_1 < v_1< u_2<v_2$), the R\'enyi entropies encode the full data of the CFT because $\Tr \rho_A^n $ is obtained as a four-point function of twist fields \cite{cct-09, cct-11}. 
By global conformal invariance we have that
\bea
\label{renyi cft 2int}
{{\rm Tr}} \rho_A^n 
&=&
\langle  \mathcal{T}_n(u_1)  \bar{\mathcal{T}}_n(v_1)
\mathcal{T}_n(u_2)  \bar{\mathcal{T}}_n(v_2)  \rangle 
\\
\label{F2n def}
\rule{0pt}{.8cm}
&=&
c_n^2\left[\frac{(u_2-u_1)(v_2-v_1)}{(v_1-u_1)(v_2-u_2)(u_2-v_1)(v_2-u_1)}\right]^{2\Delta_n} \mathcal{F}_{2,n}(x)
\eea
where the four-point ratio reads
\be
\label{x ratio def}
x = \frac{(u_1-v_1)(u_2-v_2)}{(u_1-u_2)(v_1-v_2)} \,,
\ee
and $x\in (0,1)$. 
Since ${{\rm Tr}} \rho_A =1$ holds, $\mathcal{F}_{2,1}(x)=1$ identically.
The function $\mathcal{F}_{2,n}(x)$ depends on the details of the model and therefore it must be computed case by case. 
From (\ref{twist 2point func}) and (\ref{renyi cft 2int}), one gets that (\ref{RenyiMI}) for a CFT is given by
\be
\label{RenyiMI cft}
I_{A_1,A_2}^{(n)}
=
- \, \frac{(n+1) c}{6n}\,\log(1-x)
+
\tilde{I}_n(x)\,,
\qquad
\tilde{I}_n(x) \equiv \frac{1}{n-1}\log [ \mathcal{F}_{2,n}(x) ]\,.
\ee
Since the mutual information $I_{A_1,A_2}$ is the limit $n \to 1$ of (\ref{RenyiMI cft}), as stated in (\ref{MIdef}), it is the function of $x$ given by
\be
\label{replica limit MI cft}
I_{A_1,A_2} 
=
- \, \frac{c}{3}\,\log(1-x)
+ 
\tilde{I}_1(x) \,,
\qquad
\tilde{I}_1(x) \equiv
\partial_n \mathcal{F}_{2,n}(x) \big|_{n=1}\,.
\ee

The explicit expression of $\mathcal{F}_{2,n}(x) $ is known for some simple models like the free compact boson and the Ising model.
In these cases $\mathcal{F}_{2,n}(x) $ is written in terms  of the Riemann theta function, which is defined as follows  \cite{theta books}
\be
\label{riemann theta def}
\Theta[\boldsymbol{e}](\boldsymbol{z}|\Omega) 
\equiv
\sum _{\boldsymbol{m}\;\in\;\mathbb{Z}^p}\exp
\left[\, \textrm{i} 
(\boldsymbol{m}+\boldsymbol{\varepsilon})^{\textrm{t}}
\cdot\Omega\cdot(\boldsymbol{m}+\boldsymbol{\varepsilon})
+2\pi \textrm{i} 
(\boldsymbol{m}+\boldsymbol{\varepsilon})^{\textrm{t}}\cdot (\boldsymbol{z}+\boldsymbol{\delta})\right]\,,
\ee
where $\Omega$ is a $p \times p$ symmetric complex matrix with positive immaginary part and $\boldsymbol{z} \in \mathbb{C}^p/(\mathbb{Z}^p+\Omega \mathbb{Z}^p)$ is a complex $p$ dimensional vector.
The vector $\boldsymbol{e}^{\textrm{t}} \equiv (\boldsymbol{\varepsilon^{\textrm{t}}} , \boldsymbol{\delta}^{\textrm{t}})$ is the characteristic of the Riemann theta function (\ref{riemann theta def}), being $\boldsymbol{\varepsilon}$ and $\boldsymbol{\delta}$ two $p$ dimensional vectors whose elements are either $0$ or $1/2$. 
The characteristic provides the parity of (\ref{riemann theta def}) as function of $\boldsymbol{z}$, which is the same one of the integer number $4\boldsymbol{\varepsilon}^{\textrm{t}} \cdot \boldsymbol{\delta} $, indeed
\be
\Theta[\boldsymbol{e} ]( - \boldsymbol{z} | \Omega ) 
=
(-1)^{4\boldsymbol{\varepsilon} \cdot \boldsymbol{\delta} }\,
\Theta[\boldsymbol{e} ](\boldsymbol{z} | \Omega )\,. 
\ee
It is not difficult to realize that there are $2^{p-1}(2^p+1)$ even characteristics and $2^{p-1}(2^p-1)$ odd ones.
Since in this paper we always deal with $\boldsymbol{z}=\boldsymbol{0}$, we find it convenient to lighten the formulas by introducing the notation $\Theta[\boldsymbol{e}](\Omega) \equiv \Theta[\boldsymbol{e}](\boldsymbol{0}|\Omega) $ and $\Theta(\Omega)  \equiv \Theta[\boldsymbol{0}](\Omega) $.
The Riemann theta functions throughout this paper have been evaluated by using \emph{Mathematica} through the built-in function \emph{SiegelTheta}.

As a first example, we consider the free boson compactified on a circle of radius $r$, which has $c=1$. 
The corresponding $\mathcal{F}_{2,n}(x)$ for any integer $n\geqslant 2$ is given by  \cite{cct-09}
\be
\label{F 2int compact}
\mathcal{F}_{2,n}(x)=
\frac{\Theta(\eta \tau_2 ) \, \Theta(\tau_2/\eta)}{\Theta(\tau_2)^2}\,,
\ee
where $\eta \propto r^2$ and $\tau_2= \tau_2(x)$ is the $(n-1) \times (n-1)$ purely imaginary period matrix of the Riemann surface which underlies the computation of ${{\rm Tr}} \rho_A^n$, whose elements read
\be
\label{tau2 def}
(\tau_2)_{ij}
\,\equiv\,
\textrm{i}\,\frac{2}{n}\sum_{k=1}^{n-1}\sin(\pi k/n)\, \frac{F_{k/n}(1-x)}{F_{k/n}(x)}
\,\cos[2\pi (i-j) k/n ] \,,
\ee
where $F_{s}(x) \equiv\,  _2F_1(s,1-s;1;x)$, being $_2F_1$ the hypergeometric function.
Notice that $\mathcal{F}_{2,n}(0)=1$.
Moreover, $\mathcal{F}_{2,n}(x)$ is invariant under $\eta\rightarrow 1/\eta$ and $ x \rightarrow 1-x$ separately.
The latter symmetry is related to the well known property $S_A = S_B$ of the entanglement entropy for pure states in the case of $A$ made by two disjoint intervals. 
It is worth remarking that (\ref{F 2int compact}) holds for $x\in (0,1)$. 
Indeed, when $x \in \mathbb{C}$ and $x\notin (0,1)$ the corresponding expression is slightly more complicated \cite{cct-neg-long} and it enters in the computation of the logarithmic negativity for the compact boson.

In order to find the analytic expression of the mutual information for the compact boson, one has to compute $\tilde{I}_1(x)$ in (\ref{replica limit MI cft}) with $\mathcal{F}_{2,n}(x)$ given by (\ref{F 2int compact}). 
Since performing this analytic computation is still an open problem, we employ the numerical extrapolation method suggested by \cite{ahjk-14} (see \S\ref{RIsection}) to get a result that can be compared with the numerical data found in \cite{fps-08} from the XXZ spin chain.

Before entering in the numerical analysis, it is worth discussing the decompactification regime, which can be addressed analytically.
The non compact boson corresponds to the regime $\eta \gg 1$ (or $\eta \ll 1$ because of the symmetry $\eta \leftrightarrow 1/\eta$) in the above expressions. 
In \cite{cct-09} it has been found that, for $\eta \ll 1$, the terms $\tilde{I}_1(x)$ in (\ref{replica limit MI cft}) becomes 
\be
\label{smalleta}
\tilde{I}_1(x)\big|_{\eta \ll 1} 
=
-\frac{1}{2}\log \eta + 
\frac{D(x) + D(1-x) }{2}\,,
\qquad
D(x) \equiv  
- \int_{-\textrm{i}\infty}^{\textrm{i}\infty} \frac{dz}{\textrm{i}}
\,\frac{\pi z}{\sin(\pi z)} \log[F_z(x)]\,.
\ee

The Hamiltonian of the periodic XXZ spin $1/2$ chain in a magnetic field $h$ reads
\cite{giamarchi-book}
\be
\label{XXZ spinchain}
H_{\textrm{\tiny XXZ}} \equiv
\sum_{j=1}^L \big(
S^x_j S^x_{j+1} + S^y_j S^y_{j+1} + \Delta\, S^z_j S^z_{j+1}
\big)
- h \sum_{j=1}^L S^z_j\,,
\ee
where $S^a_j = \sigma^a_j / 2$, being $\sigma^a_j$ the standard Pauli matrices acting on the spin at the $j$-th site. 
The chain has $L$ sites and $\Delta$ is the anisotropy. 
The mutual information for this lattice model has been computed in \cite{fps-08} by direct diagonalization for $L \leqslant 30$.
When $h=0$ and $-1 < \Delta \leqslant 1$ the model in the continuum is described by the $c=1$ compact boson with $\eta = 1 -(1/\pi) \arccos \Delta $, while for $h \neq 0$ an explicit formula providing $\eta$ does not exist and therefore it must be found numerically.
The CFT formulas reviewed above can be applied also to the case of a finite system of length $L$ with periodic boundary conditions by employing a conformal mapping from the cylinder to the plane. 
As final result, the CFT formulas for this case are obtained by considering the expressions for the infinite line and replacing any length $\ell_i $ with the corresponding chord length $(L/\pi) \sin(\pi\ell_i/L)$ \cite{cc-04}.

\begin{figure}[t]
\vspace{.0cm}
\hspace{-.8cm}
\includegraphics[width=17.5cm]{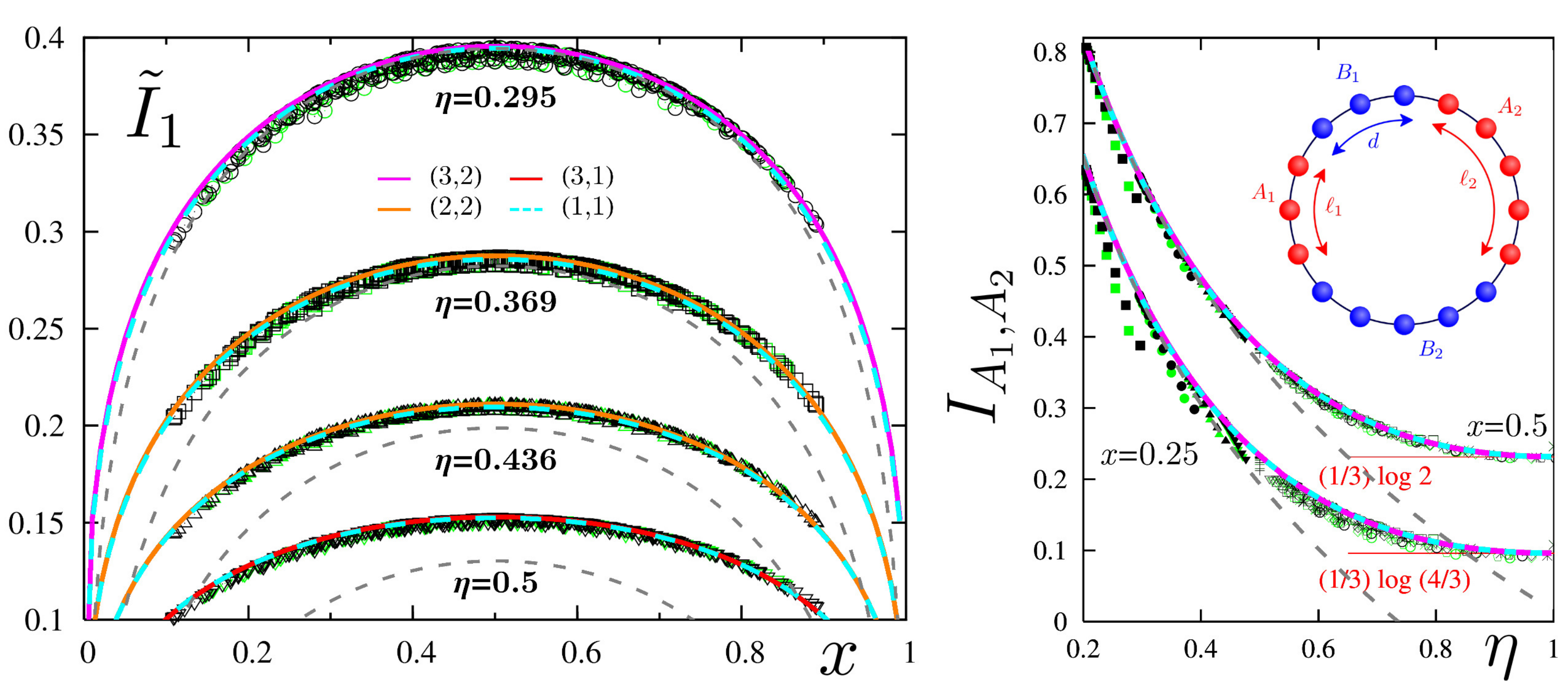}
\vspace{-.5cm}
\caption{
Mutual information for the XXZ model. 
The data points are extracted from \cite{fps-08} and the coloured curves are obtained from the rational interpolations of the analytic expressions (\ref{RenyiMI cft}) and (\ref{F 2int compact}) for the compact boson with the values of $(p,q)$ indicated in the left panel. The dashed grey lines correspond to the decompactification regime, where the analytic continuation (\ref{smalleta}) is known.
Left: $\tilde I_1$, defined in (\ref{replica limit MI cft}), as function of $x$ for various values of $\eta$. Right: the mutual information $I_{A_1,A_2}$ as function of $\eta$ for two fixed values of $x$.
}
\label{fig:RI_FPS}
\end{figure}

Let us consider the mutual information of the compactified boson as first example of our extrapolation method. 
For any fixed value of $x$, we have that $I_{A_1,A_2}^{(n)}$ are given analytically by (\ref{RenyiMI cft}) and (\ref{F 2int compact}) for any positive integer $n \geqslant 2$, while the corresponding analytic continuation to $n=1$ is estimated by performing a numerical extrapolation of the known data through a rational function.
The latter one is characterized by two positive integer parameters $p$ and $q$, which are  the degrees of the numerator and of the denominator respectively.
As explained in \S\ref{RIsection}, to perform a rational interpolation characterized by the pair $(p,q)$ we need at least $p+q+1$ known data.
An important technical difficulty that one encounters is the evaluation of the Riemann theta functions for large genus period matrices, i.e. for high values of $n$.
Given the computational resources at our disposal, we were able to compute Riemann theta functions containing matrices whose size is at most $12$.
For the compactified boson this corresponds to $n_{\text{max}}=11$ and therefore
$p+q+1\leqslant 10$.

In Fig.\,\ref{fig:RI_FPS} we compared our numerical extrapolations of the analytic expressions of \cite{cct-09} with the numerical data for the XXZ spin chain computed in \cite{fps-08} by exact diagonalization, finding a very good agreement.
In the left panel $\tilde I_1$ is shown as function of the four-point ratio $x$ for different values of the parameter $\eta$, while in the right panel the mutual information $I_{A_1,A_2}$ is shown as function of $\eta$ for the two fixed  configurations of intervals given by $\ell_1=\ell_2=d_1=d_2=L/4$ ($x=0.5$) and $2\ell_1=2\ell_2=d_1=d_2=L/3$ ($x=0.25$), being $L$ the total length of the periodic system.
All the rational interpolations in the figure exhibit a good agreement with the numerical data, despite the low values of $p$ and $q$. 
Increasing these parameters, a better approximation is expected but the result is already stable for these values and 
we provided two rational interpolations for each curve as a check.
Some rational interpolations may display some spurious bahaviour in some regimes of $x$. As discussed in detail in \S\ref{RIsection}, this possibility increases with $q$.
These results have been discarded and  we showed only rational interpolations which are well-behaved in the whole domain $x\in (0,1)$.
Notice that rational interpolations that are well-behaved for some $\eta$ and $x$ could display some bad behaviour changing them. 
Thus, the values of $(p,q)$ must be chosen case by case. 
In Fig.\,\ref{fig:RI_FPS} the dashed grey lines are obtained from the analytic continuation (\ref{smalleta}) found in \cite{cct-09}, which corresponds to the decompactification regime and therefore it reproduces the numerical data from the XXZ chain and from the rational interpolations only for small $\eta$, as expected.

Another important case where the R\'enyi entropies of two disjoint intervals have been found analytically is the Ising model \cite{cct-11}. 
The Hamiltonian of the one dimensional spin chain defining the Ising model in a transverse field is 
\be
\label{Ham Ising}
H_{\textrm{\tiny Ising}} \equiv
- \sum_{j=1}^L \big(
\sigma^x_j \sigma^x_{j+1}  + h\, \sigma^z_j 
\big)\,,
\ee
where periodic boundary conditions are imposed. 
This model has a quantum critical point at $h=1$ and in the continuum it is a free Majorana fermion with central charge $c=1/2$.
The R\'enyi entropies for two disjoint intervals on the spin chain (\ref{Ham Ising}) have been studied in \cite{atc-10} through a Tree Tensor Network algorithm \cite{TTNalgo} and in \cite{fc-10} through the exact solution of the model in terms of free Majorana fermions. 
The former method allowed to find also the mutual information.

\begin{figure}[t]
\vspace{.2cm}
\hspace{.8cm}
\includegraphics[width=14cm]{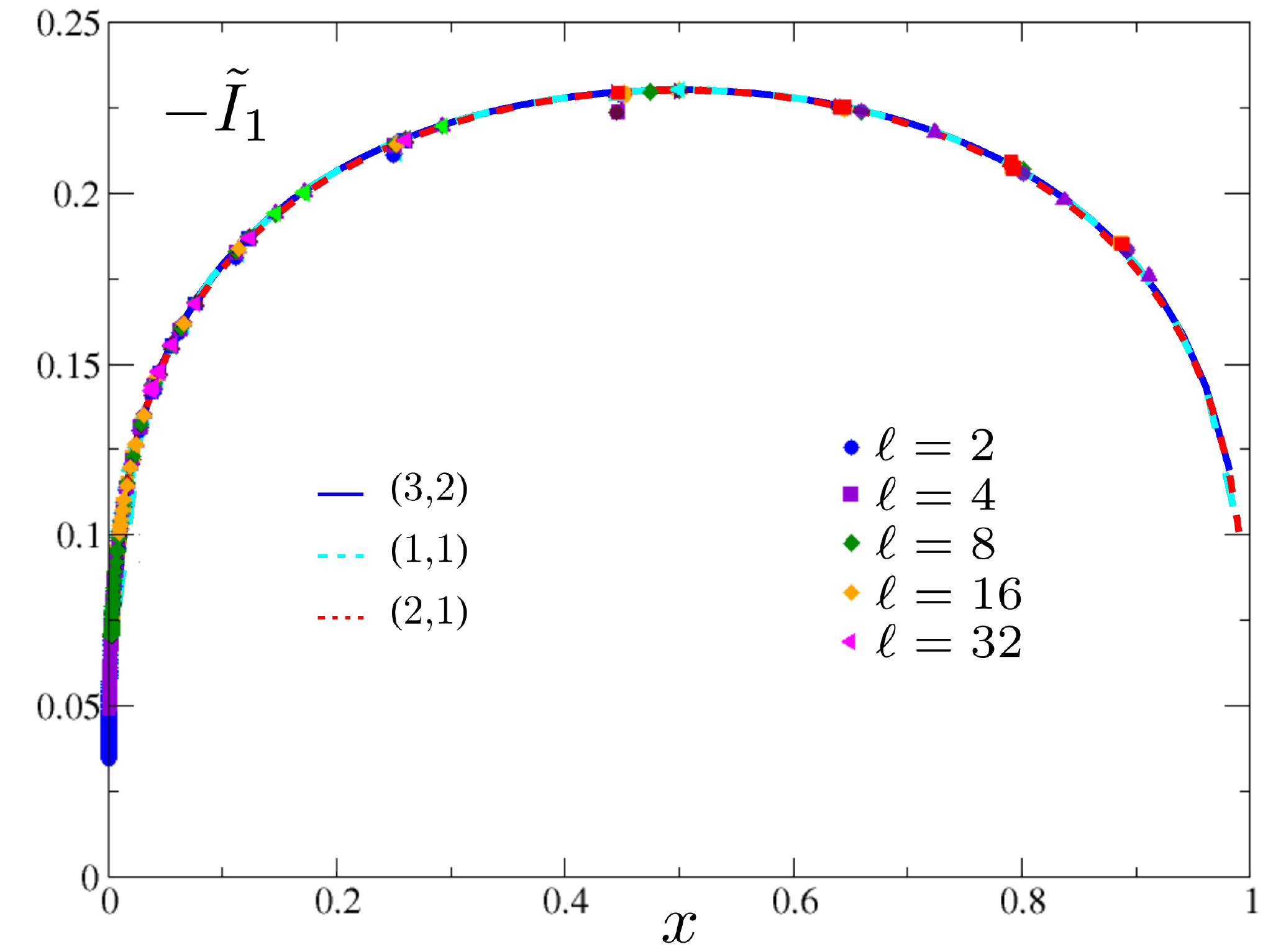}
\vspace{.2cm}
\caption{
Extrapolations for $-\tilde I_1$, defined in (\ref{replica limit MI cft}), as function of $x$ for the Ising model. The data points are extracted from \cite{atc-10} while the coloured curves are obtained through the rational interpolations with $(p,q)$ indicated.  
}
\label{fig:RI_ATC}
\end{figure}

As for the R\'enyi entropies for two disjoint intervals in corresponding CFT, by employing known results about bosonization on higher genus Riemann surfaces for $c=1$ models \cite{bosonization higher genus}, the expression of $\mathcal{F}_{2,n}(x)$ for the Ising model can be written in terms of Riemann theta functions (\ref{riemann theta def}) evaluated for the period matrix  $\tau_2$ in (\ref{tau2 def}).
In particular, $\Tr \rho_{A_1 \cup A_2}^n$ for the Ising model is given by (\ref{F2n def}) with $c=1/2$ and \cite{cct-11}
\be
\label{FIsing N=2}
\mathcal{F}_{2,n}(x)
=
\frac{\sum_{\boldsymbol{e}}\left|\Theta[\boldsymbol{e}](\tau_2)\right|}{2^{n-1}\left| \Theta(\tau_2)\right|}\,,
\ee
where the sum is performed over all the possible characteristics 
$\boldsymbol{e}^{\textrm{t}} \equiv (\boldsymbol{\varepsilon^{\textrm{t}}} , \boldsymbol{\delta}^{\textrm{t}})$, being $\boldsymbol{\varepsilon}$ and $\boldsymbol{\delta}$ two $n-1$ dimensional vectors whose elements are either $0$ or $1/2$.
Let us remark that in the sum (\ref{FIsing N=2}) only the $2^{n-2}(2^{n-1}+1)$ even characteristics occur. 
Thus, the mutual information for the Ising model is (\ref{replica limit MI cft}) with $\mathcal{F}_{2,n}(x)$ given by (\ref{FIsing N=2}).
Similarly to the case of the compact boson, also for the Ising model we are not able to compute $\tilde{I}_1(x) $ analytically and therefore we perform a numerical extrapolation through the rational interpolation method described in \S\ref{RIsection}.

In Fig.\,\ref{fig:RI_ATC} we show $- \tilde{I}_1(x) $ as function of $x\in(0,1)$, which can be found by considering two disjoint intervals of equal length, and compare the numerical data obtained in \cite{atc-10} with the curve found through the numerical extrapolation of the corresponding formula containing (\ref{FIsing N=2}) through rational interpolations.
Since (\ref{FIsing N=2}) contains Riemann theta functions, we cannot consider high values for $n$, like for the compact boson. 
Moreover, in this case one faces an additional complication with respect to the compact boson because in (\ref{FIsing N=2}) the sum over all the even characteristics occurs and the number of terms in the sum grows exponentially with $n$.  
Given our computational power, we have computed the R\'enyi entropies up to $n=7$
and in Fig.\,\ref{fig:RI_ATC} we show the rational interpolations found by choosing three different pairs $(p,q)$ which are well-behaved among the available ones. 
Since the curves coincide, the final result is quite stable and, moreover, the agreement with the numerical data found in \cite{atc-10} through the Tree Tensor Network is very good.


\section{Three disjoint intervals}
\label{sec tripartite}

In this section we partially extend the analysis done in \S\ref{sec MI} by considering the case of three disjoint intervals. 
After a brief review of the analytic results known for a generic number $N$ of disjoint intervals, we focus on $N=3$ and perform some numerical extrapolations for the non compact boson and for the Ising model.

Given a the spatial subsystem $A = \cup_{i=1}^N A_i$ made by the union of the $N$ disjoint intervals $A_1 = [u_1, v_1]$, $\dots$, $A_N = [u_N, v_N]$, a generalization of (\ref{RenyiMI}) to $N \geqslant 2$ reads \cite{ctt-14}
\be
\label{renyi MI Nint}
I^{(n)}_{A_1,\cdots, A_N}\equiv \frac{(-1)^N}{n-1}\log R_{N,n} \,,
\qquad 
R_{N,n}\equiv 
\prod_{p=1}^N\prod_{\sigma_{N,p}}\left(\Tr \rho^n_{\sigma_{N,p}}\right)^{(-1)^{N-p}} \,,
\ee
where $\sigma_{N,p}$ denotes the union of a generic choice of $1\leqslant p \leqslant N$ intervals among the $N$ ones. 
It is straightforward to observe that the analytic continuation $n \to 1$ of (\ref{renyi MI Nint}), i.e.
\be
\label{N mutual info def}
I_{A_1 , \dots , A_N} \equiv \lim_{n \rightarrow 1} I_{A_1 , \dots , A_N}^{(n)} \,,
\ee 
provides a natural generalization to $N \geqslant 2$ of the mutual information (\ref{MIdef}).
We find it useful to normalise the quantities introduced in (\ref{renyi MI Nint}) and (\ref{N mutual info def}) by themselves evaluated for some fixed configuration of intervals, namely
\be
\label{RNn norm}
R^{\textrm{\tiny{norm}}}_{N,n} 
\equiv \frac{R_{N,n}}{
R_{N,n}\big|_{\textrm{\tiny{fixed}}}}\,,
\qquad
I^{\textrm{\tiny{sub}}}_{N} 
\equiv 
I_{N}  -  I_{N}\big|_{\textrm{\tiny{fixed}}}
=
\lim_{n\to 1} R^{\textrm{\tiny{norm}}}_{N,n} \,,
\ee
where we have adopted the shorthand notation $I_{N} \equiv I_{A_1 , \dots , A_N} $.

In two dimensional CFTs, the expression of $\Tr \rho_A^n$ for $N$ disjoint intervals can be written as a $2N$-point function of twist fields \cite{cc-04, cc-rev}.
Similarly to the two intervals case, the global conformal invariance cannot fix the dependence on $u_i$ and $v_i$.
In particular, given the endpoints $u_1<v_1<\dots<u_N<v_N$, one can employ the following conformal map
\be
w_N(z)=\frac{(u_1-z)(u_N-v_N)}{(u_1-u_N)(z-v_N)}\,,
\ee
which sends $u_1\rightarrow0$, $u_N\rightarrow1$ and $v_N\rightarrow\infty$. 
The remaining endpoints are mapped into the $2N-3$ four-point ratios $x_1=w_N(v_1),\, x_2= w_N(u_2),\, x_3=w_N(v_2), \dots, x_{2N-3}=w_N(v_{N-1})$ which are invariant under $SL(2,\mathbb{C})$. 
Notice that $x_j \in \mathbb{R}$ and the order is preserved, namely $0<x_1<x_2<\dots<x_{2N-3}<1$. 

The global conformal invariance allows us to write $\Tr \rho_A^n$ for $N$ disjoint intervals as follows \cite{cc-rev}
\be
\label{Tr rhoA Nint cft}
\Tr \rho_A^n
=
\langle \prod_{i=1}^N \mathcal{T}_n(u_i)\mathcal{\bar{T}}_n(v_i) \rangle
=
c_n^N\left|
\frac{\prod_{i< j}(u_j-u_i)(v_j-v_i)}{\prod_{i,j}(v_j-u_i)}
\right|^{2\Delta_n}
\mathcal{F}_{N,n}(\boldsymbol{x})\,,
\ee
where $i,j=1,\dots,N$, the scaling dimension $\Delta_n$ is given in (\ref{twist 2point func}) and $\boldsymbol{x}$ is the vector whose elements are the $2N-3$ four-point ratios introduced above.
It is worth remarking that $ \mathcal{F}_{N,n}(\boldsymbol{x})$ encodes the full operator content of the model and therefore its computation depends on the features of the model.
From (\ref{renyi MI Nint}) and (\ref{Tr rhoA Nint cft}), one finds that $R_{N,n}$ and $R^{\textrm{\tiny{norm}}}_{N,n}$ in CFT become respectively \cite{ctt-14}
\be
\label{RNn cft}
R_{N,n}(\boldsymbol{x}) = 
\prod_{p\,=\,2}^{N}\, \prod_{\sigma_{N,p}} 
\big[  \mathcal{F}_{p,n}(\boldsymbol{x}^{\sigma_{N,p}})\big]^{(-1)^{N-p}} \,,
\qquad
R^{\textrm{\tiny{norm}}}_{N,n}(\boldsymbol{x}) 
= \frac{R_{N,n}(\boldsymbol{x})}{
R_{N,n}(\boldsymbol{x}_{\textrm{\tiny{fixed}}})}\,,
\ee
where $\boldsymbol{x}^{\sigma_{N,p}}$  is the vector made by the $2p-3$ four-point ratios obtained with the endpoints of the $p$ intervals selected by $\sigma_{N,p}$.

The function $\mathcal{F}_{N,n}(\boldsymbol{x})$ for the compactified boson has been  studied in \cite{ctt-14} by generalizing the construction of \cite{cct-09} and, again, it is written in terms of the Riemann theta function (\ref{riemann theta def}). 
For $N > 2$ disjoint intervals the Riemann surface occurring in the computation of $\Tr \rho_A^n$ has genus $g=(N-1)(n-1)$.
The corresponding $g\times g$ period matrix $\tau_N = \mathcal{R}+ \textrm{i}\,\mathcal{I}$, which is symmetric and complex with positive imaginary part, is complicated and, since we do not find instructive to report it here, we refer to \cite{ctt-14} for any detail about it.
The expression of $\mathcal{F}_{N,n}(\boldsymbol{x})$ for the compactified boson reads \cite{bosonization higher genus, ctt-14} 
\be
\label{F Nint compact boson}
\mathcal{F}_{N,n}(\boldsymbol{x})
= 
\frac{\Theta(T_{\eta})}{|\Theta(\tau_N)|^2}\,,
\qquad
T_{\eta} \equiv
\begin{pmatrix}
\,\textrm{i}\,\eta\,\mathcal{I} & \mathcal{R} \\
\mathcal{R} & \textrm{i}\,\mathcal{I}/\eta\, \\
\end{pmatrix} ,
\ee
where $\eta$ is the parameter containing the compactification radius introduced in \S\ref{sec MI}. Notice that (\ref{F Nint compact boson}) is invariant under $\eta \leftrightarrow 1/\eta$.

As done in \S\ref{sec MI} for the two intervals case, also for $N$ disjoint intervals it is interesting to consider the decompactification regime. 
When $\eta \gg 1$ the expression in (\ref{F Nint compact boson}) becomes
\be
\label{FNn dec boson}
\mathcal{F}_{N,n}^{\eta\rightarrow \infty}( \boldsymbol{x}  ) 
=
\frac{\eta^{g/2}}{ \sqrt{\textrm{det}(\mathcal{I})}\, | \Theta ( \tau_N ) |^2}
\equiv
\eta^{g/2}\,\widehat{\mathcal{F}}_{N,n}( \boldsymbol{x}  ) \,.
\ee
For computational purposes, it is important to observe that in (\ref{FNn dec boson}) the Riemann theta function is evaluated for $\tau_N$, which is $g\times g$, while for finite $\eta$, when (\ref{F Nint compact boson}) holds, the matrix occurring in the Riemann theta function is $2g\times 2g$.
This implies that for the non compact boson we can reach higher values of $n$ and therefore the corresponding numerical extrapolation is more precise. 
In the decompactification regime we can also appreciate the convenience of considering the normalization (\ref{RNn norm}). Indeed, plugging (\ref{FNn dec boson}) into (\ref{RNn cft}) one obtains an expression which is $\eta$ independent
\be
\label{rNn dec def}
\widehat{R}^{\textrm{\tiny{norm}}}_{N,n}(\boldsymbol{x}) 
\equiv
\frac{R_{N,n}^{\eta \rightarrow \infty}( \boldsymbol{x})}{R_{N,n}^{\eta \rightarrow \infty}( \boldsymbol{x}_{\textrm{\tiny{fixed}}})}
=
\prod_{p\,=\,2}^{N}\, \prod_{\sigma_{N,p}} 
\left[
\frac{\widehat{\mathcal{F}}_{p,n}(\boldsymbol{x}^{\sigma_{N,p}})}{\widehat{\mathcal{F}}_{p,n}(\boldsymbol{x}^{\sigma_{N,p}}_{\textrm{\tiny{fixed}}})}
\right]^{(-1)^{N-p}} .
\ee

As for the Ising model, since the results of \cite{bosonization higher genus} about the bosonization on higher genus Riemann surfaces for $c=1$ models hold for a generic genus, we can straightforwardly write the generalization to $N \geqslant 2$ of the $N=2$ formula (\ref{FIsing N=2}).
Indeed, given the period matrix $\tau_N$ employed for the compact boson in (\ref{F Nint compact boson}), we have that $\Tr \rho_{A}^n$ for the Ising model is (\ref{Tr rhoA Nint cft}) with $c=1/2$ and \cite{bosonization higher genus, ctt-14}
\be
\label{FIMGeneralN}
\mathcal{F}_{N,n}(\boldsymbol{x})
= 
\frac{\sum_{\boldsymbol{e}}|\Theta[\boldsymbol{e}](\tau_N)|}{
2^g\,|\Theta( \tau_N)|}\,.
\ee
The Riemann theta functions in this formula are evaluated for the $g \times g$ period matrix and a sum over all the characteristics occurs. 
It is worth remarking that the Riemann theta functions in (\ref{FIMGeneralN}) with odd characteristics vanish and therefore the sum contains $2^{g-1}(2^g+1)$ terms.
In \cite{ctt-14} the formula (\ref{FIMGeneralN}) has been checked numerically on the lattice for $n=2$, various $N$ and different configurations of intervals by employing the Matrix Product States.
To our knowledge, numerical results for $I_{N} $ with $N\geqslant 3$ are not available in the literature for the critical Ising chain in transverse field.

In this paper, for simplicity, we consider only $N=3$ disjoint intervals and therefore let us  specify some of the formulas given above to this case. 
The generalization of the mutual information to the case of three disjoint intervals is given by
\be
\label{I3 def}
I_{A_1,A_2,A_3}
\equiv
S_{A_1}+S_{A_2}+S_{A_3}
-S_{A_1\cup A_2}-S_{A_1\cup A_3}-S_{A_2\cup A_3}
+ S_{A_1\cup A_2\cup A_3}
= 
\lim_{n \to 1} 
I^{(n)}_{A_1,A_2,A_3}\,,
\ee
where $I^{(n)}_{A_1,A_2,A_3}$ can be written by specifying the expressions in (\ref{renyi MI Nint}) to $N=3$, namely
\be
\label{I3n def}
I^{(n)}_{A_1,A_2,A_3}
\equiv\,
 \frac{\log (R_{3,n}) }{1-n}
\,=\,
S^{(n)}_{A_1}+S^{(n)}_{A_2}+S^{(n)}_{A_3}-S^{(n)}_{A_1\cup A_2}-S^{(n)}_{A_1\cup A_3}-S^{(n)}_{A_2\cup A_3}+ S^{(n)}_{A_1\cup A_2\cup A_3}\,,
\ee
with
\be 
R_{3,n}
\equiv
\frac{\Tr \rho^n_{A_1 \cup A_2 \cup A_3}
\big(\Tr \rho^n_{A_1} \Tr \rho^n_{A_2} \Tr \rho^n_{A_3} \big)}{
\Tr \rho^n_{A_1 \cup A_2} \Tr \rho^n_{A_1 \cup A_3} \Tr \rho^n_{A_2\cup A_3}}\,.
\ee
Considering CFTs, when $N=3$ the vector $\boldsymbol{x}=(x_1, x_2, x_3)$ is made by three four-point ratios and (\ref{RNn cft}) becomes
\be
\label{RNn cft N=3}
R_{3,n} (\boldsymbol{x}) 
=
\frac{\mathcal{F}_{3,n}(x_1,x_2,x_3)}{
\mathcal{F}_{2,n}(\frac{x_1(x_3-x_2)}{x_2(x_3-x_1)}) 
\, \mathcal{F}_{2,n}(x_1)
\, \mathcal{F}_{2,n}(\frac{x_3-x_2}{1-x_2}) }\,,
\ee
where $\mathcal{F}_{3,n}(\boldsymbol{x})$ is (\ref{Tr rhoA Nint cft}) for $N=3$ and
$\mathcal{F}_{2,n}(x) $ has been introduced in (\ref{F2n def}).

\begin{figure}[t]
\vspace{.3cm}
\hspace{-.2cm}
\includegraphics[width=16cm]{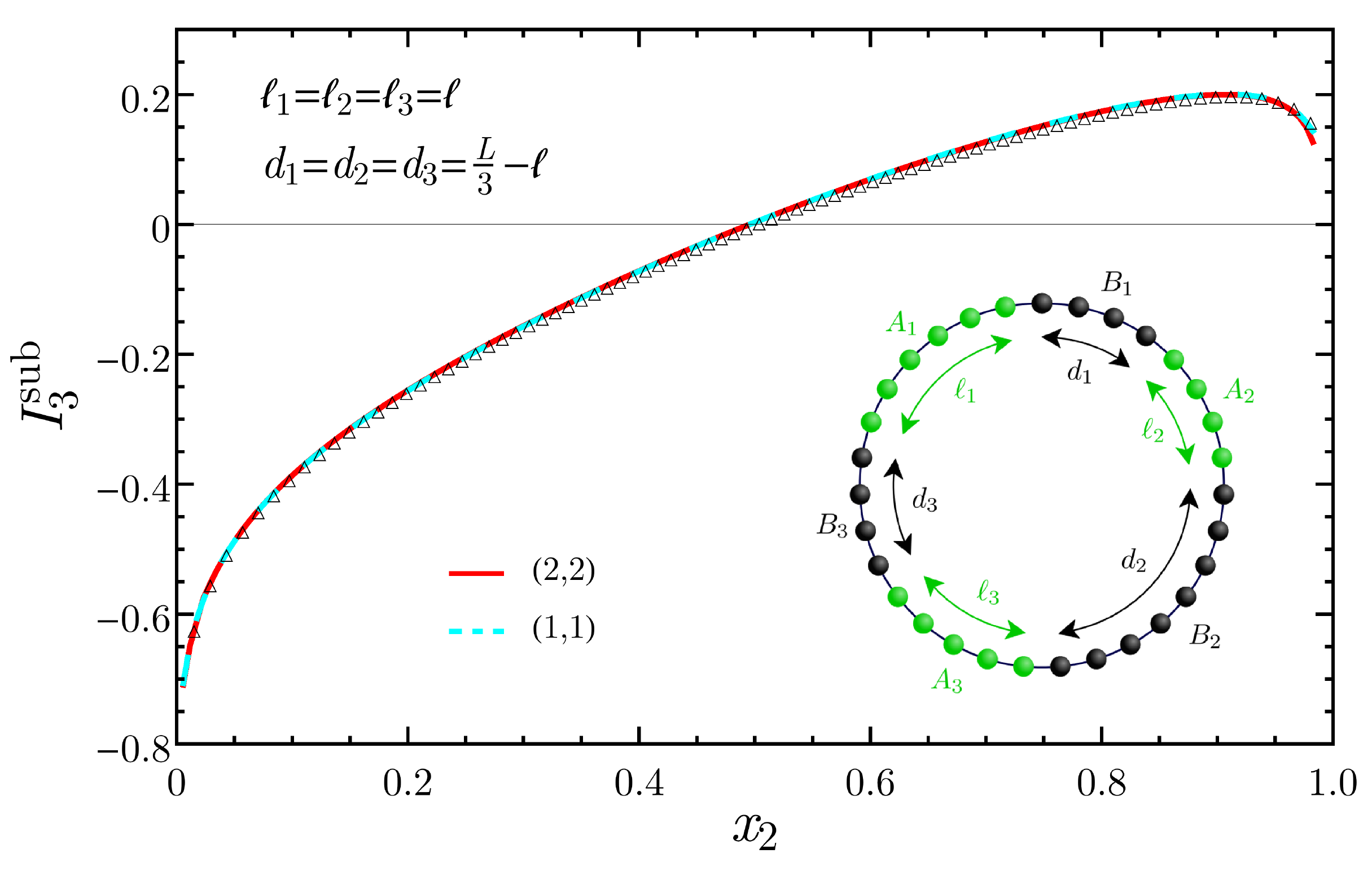}
\vspace{-.3cm}
\caption{
Extrapolations of $I^{\textrm{\tiny{sub}}}_{3}$ (see (\ref{RNn norm}) with $N=3$) as function of the four-point ratio $x_2$ for the non compact boson.
The points are the data obtained in \cite{ctt-14} from the periodic harmonic chain (\ref{HC ham}) with  $L=5000$ and $\omega L = 10^{-5}$.
The configuration chosen here is made by equal intervals separated by equal distances, while the fixed configuration normalizing $I^{\textrm{\tiny{sub}}}_{3}$ is given in the text.
The coloured lines correspond to two different extrapolations obtained through rational interpolations with $(p,q)$ indicated.
}
\label{fig:RI_I3_DecBos}
\end{figure}

\begin{figure}[t]
\vspace{.3cm}
\hspace{-.2cm}
\includegraphics[width=16cm]{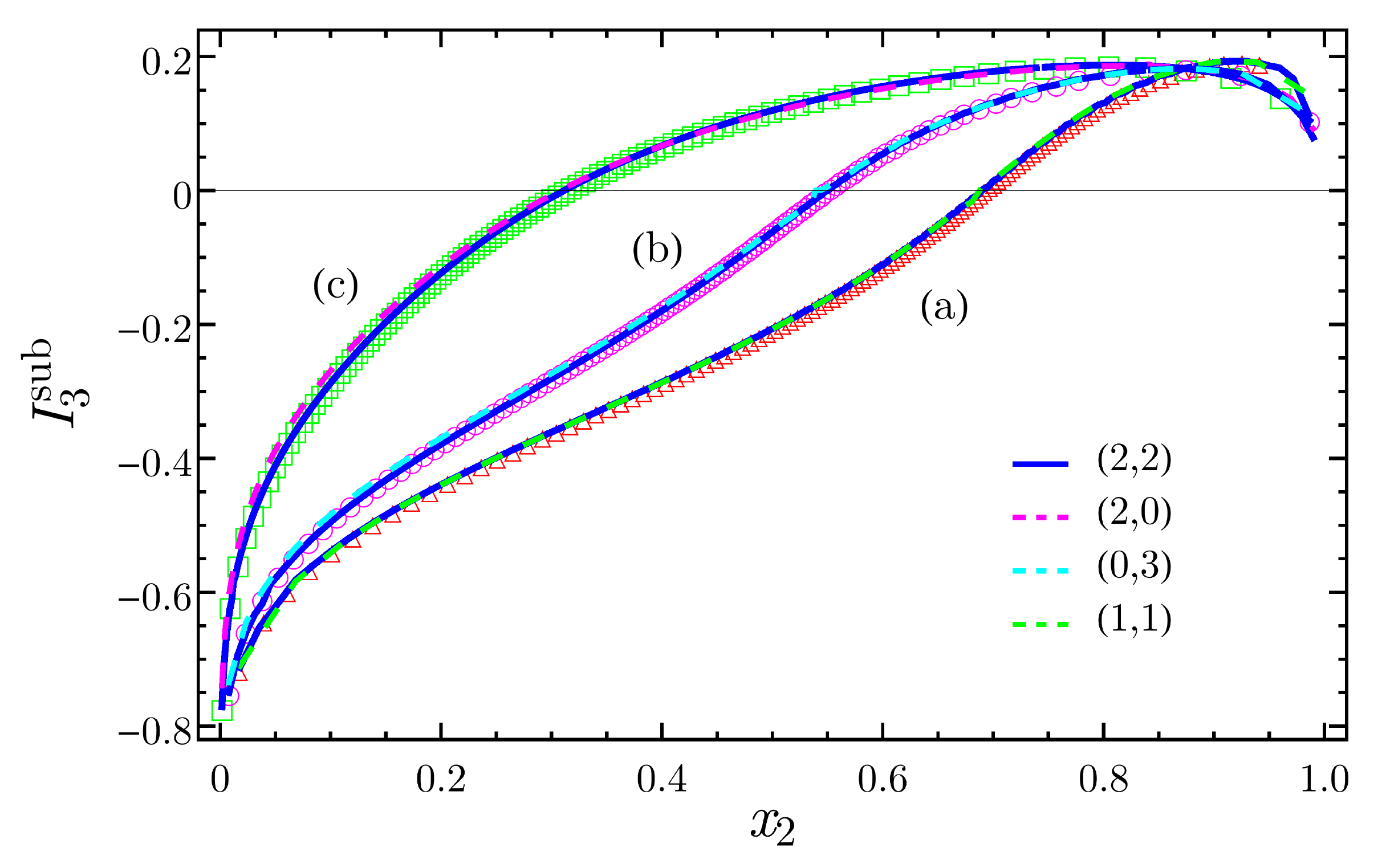}
\vspace{-.3cm}
\caption{
Extrapolations of $I^{\textrm{\tiny{sub}}}_{3}$ for the non compact boson.
The harmonic chain is the same one of Fig.\,\ref{fig:RI_I3_DecBos} while the configurations of intervals are given by (\ref{configs intervals}).
The data for the periodic harmonic chain have been extracted from \cite{ctt-14}.
}
\label{fig:RI_I3_DecBos_All}
\end{figure}

The non compact  boson is the CFT describing the massless harmonic chain in the continuum. 
The Hamiltonian of the harmonic chain with $L$ lattice sites and with nearest neighbour interaction reads
\be
\label{HC ham}
H = \sum_{n=0}^{L-1} \left(
\frac{1}{2M}\,p_n^2+\frac{M\omega^2}{2}\,q_n^2 +\frac{K}{2}(q_{n+1} -q_n)^2
\right),
\ee
where periodic boundary conditions are imposed. 
Rewriting (\ref{HC ham}) in terms of $a\equiv \sqrt{M/K}$ and $\omega$ through a canonical transformation, one can observe that it provides the lattice discretization of the free boson with mass $\omega$ and lattice spacing $a$.
Thus, the continuum limit  of the $\omega=0$ case is the decompactified boson discussed above. 
The method to compute R\'enyi entropies for the lattice model (\ref{HC ham}) is well known \cite{entHC} and $\Tr \rho_A^n $ can be found from the correlators $\langle q_r q_s \rangle $ and $\langle p_r p_s \rangle $.
Let us recall that setting $\omega$ to zero leads to a divergent expression for $\langle q_r q_s \rangle $ because of the zero mode occurring for periodic boundary conditions.
In \cite{ctt-14} the method discussed in \cite{entHC} has been applied to perform various checks of the CFT formulas for the non compact boson at fixed $n$.
Moreover, also $I^{\textrm{\tiny{sub}}}_{N} $ has been found from the harmonic chain data, but a comparison with the analytic results has not been done because the analytic continuation of the corresponding R\'enyi entropies is not known yet. 
Indeed, the Riemann theta function occurs in (\ref{FNn dec boson}) and its analytic continuation in $n$ is still an open problem. 
As for the values of $\omega$, in \cite{ctt-14} it has been checked that $\omega L=10^{-5}$ is small enough to capture the CFT regime through the periodic harmonic chain. The numerical data for the periodic harmonic chain have been found by setting $M=K=1$ and $\omega L=10^{-5}$ in (\ref{HC ham}). The same quantities evaluated for $\omega L=10^{-3}$ turned out to be indistinguishable.

In the remaining part of this section we focus on the case of three disjoint intervals and perform some numerical extrapolations of the analytic results reviewed above to $n = 1$ through rational interpolations, comparing them with the corresponding numerical data from the lattice models, whenever they are available. 

In Figs.\,\ref{fig:RI_I3_DecBos} and \ref{fig:RI_I3_DecBos_All} we consider $I^{\textrm{\tiny{sub}}}_{3}$ (see (\ref{RNn norm})) for the decompactified boson, comparing the results obtained for the periodic harmonic chain with the numerical extrapolations found for the corresponding configurations of intervals obtained through the rational interpolation (see \S\ref{RIsection}). 
The dots are numerical data obtained in \cite{ctt-14} from the periodic harmonic chain given by (\ref{HC ham}) with $L=5000$ and different sets of data correspond to different configurations of the three intervals.
In particular, referring to the inset of Fig.\,\ref{fig:RI_I3_DecBos} for the notation, the configuration considered in Fig.\,\ref{fig:RI_I3_DecBos} is the one where all intervals are equal $\ell_1=\ell_2=\ell_3$ and they are placed at the same distance $d_1=d_2=d_3=L/3-\ell$.
Varying the length $\ell$ of the intervals, one finds the result, which is plotted as function of the four-point ratio $x_2$.
In Fig.~\ref{fig:RI_I3_DecBos_All}, the data are labeled according to the following configurations of the three intervals:
\be
\label{configs intervals}
\begin{array}{ccl}
\textrm{(a)}
& &
\textrm{$\ell_i=\lambda_i\ell$, $d_i=(L-\sum_{i=1}^3\ell_i)/3$ with $\lambda_1=1$, $\lambda_2=2$, $\lambda_3=8$;}
\\
\textrm{(b)}
& &
\textrm{\phantom{$\ell_i=\lambda_i\ell$, $d_i=(L-\sum_{i=1}^3\ell_i)/3$} 
with $\lambda_1=1$, $\lambda_2=11$, $\lambda_3=11$;}
\\
\textrm{(c)}
& &
\textrm{$\ell_i=\gamma_i\ell$, $d_i=\gamma_i d$, $d=L/(\sum_{i=1}^3\gamma_i)-\ell$ 
  with $\gamma_1=1$, $\gamma_2=3$, $\gamma_3=6$;}
\end{array}
\ee
where the parameter $\ell$ is varied and the results are plotted as functions of $x_2\in (0,1)$.
As for the fixed configuration normalizing $I^{\textrm{\tiny{sub}}}_{3} $ in (\ref{RNn norm}) we have chosen $\ell_1=\ell_2=\ell_3=d_1=d_2=\textrm{int}(L/6)$, where $\textrm{int}(\dots)$ denotes the integer part.
The coloured curves in Figs.\,\ref{fig:RI_I3_DecBos} and \ref{fig:RI_I3_DecBos_All} are the numerical extrapolations of the CFT formulas for the non compact boson (\ref{FNn dec boson}) and (\ref{rNn dec def}) through the rational interpolation method. 
For each set of data, we show two different rational interpolations which are well-behaved in order to check the stability of the result.
The differences between different well-behaved rational interpolations are very small and the agreement with the numerical data from the harmonic chain is very good, supporting the validity of the extrapolating method.
In Figs.\,\ref{fig:RI_I3_DecBos} and \ref{fig:RI_I3_DecBos_All} we have employed $2\leqslant n \leqslant 6$.
It is worth remarking at this point that the Riemann theta functions occurring in the CFT expression (\ref{RNn cft N=3}) for the non compact boson contain at most $g \times g$ matrices ($g=2(n-1)$ for $N=3$) while for the compact boson their size is at most $2g \times 2g$ (see (\ref{F Nint compact boson})). From the computational viewpoint, this is an important difference  because the higher is $n$ that can be addressed, the higher is the number of different $(p,q)$ that can be considered in the rational interpolations.
Thus, the maximum $n$ that we can deal with is related to the maximum size of the matrices in the Riemann theta functions occurring in the model. 
Nevertheless, from Figs.\,\ref{fig:RI_I3_DecBos} and \ref{fig:RI_I3_DecBos_All}  we observe that, for this case, rational interpolations with low values of $(p,q)$ are enough to capture the result expected from the lattice data.

\begin{figure}[t]
\vspace{.2cm}
\hspace{-1.6cm}
\includegraphics[width=18.5cm]{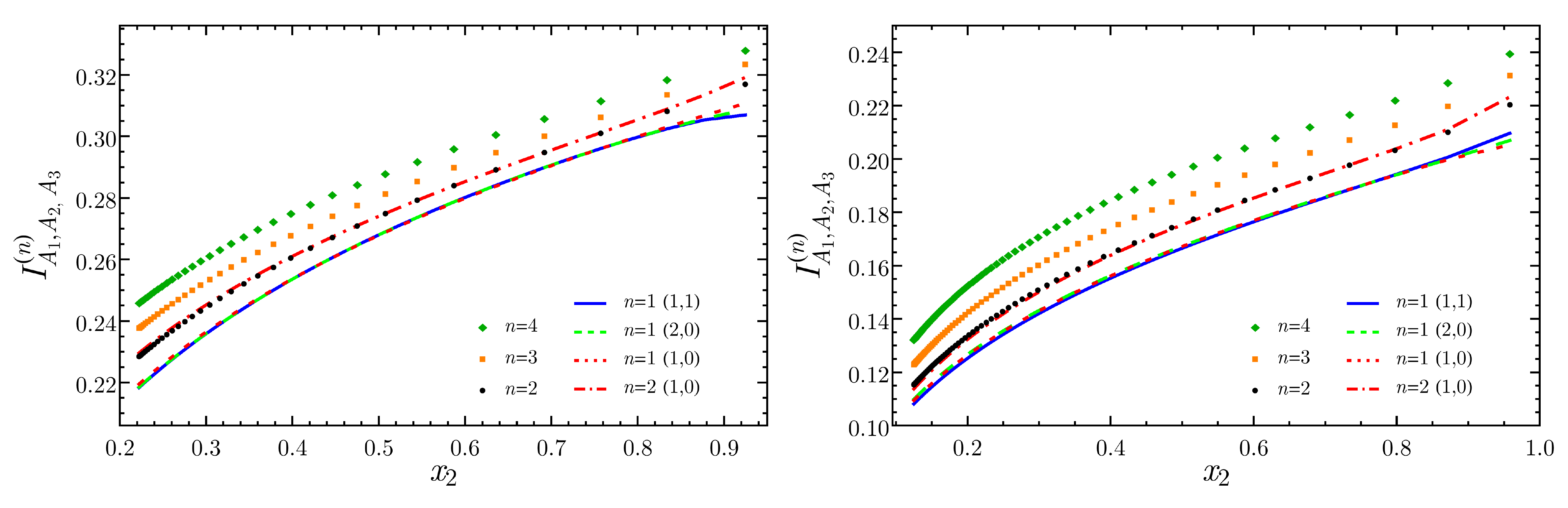}
\vspace{-.6cm}
\caption{
Extrapolations of $I_{A_1,A_2,A_3}$, defined in (\ref{I3 def}), for the Ising model.
Two configurations of intervals have been considered, namely (\ref{config 3 intervals d}) with $\alpha=0.25$ (left) and $\alpha=2$ (right). 
The dots correspond to $I^{(n)}_{A_1,A_2,A_3}$ in (\ref{I3n def}) with $n\in \{2,3,4\}$ while the lines are the extrapolations obtained through the rational interpolation method with the values of $(p,q)$ indicated. The dot-dashed line is the extrapolation to $n=2$ performed as a check of the method, while the remaining lines correspond to $I_{A_1,A_2,A_3}$.
}
\label{fig:I3Ising_both}
\end{figure}

In Fig.\,\ref{fig:I3Ising_both} we show $I_{A_1,A_2,A_3}$, defined in (\ref{I3 def}), for the Ising model.
We have considered the following configurations of three intervals specified by a parameter $\alpha$ (see the inset of Fig.\,\ref{fig:RI_I3_DecBos} for the notation) 
\be
\label{config 3 intervals d}
\begin{array}{ccl}
\textrm{(d)}
& &
\textrm{$\ell_i= \ell$, $d_1=d_2=\alpha \ell$, $d_3 = L - (3+2\alpha)\ell$\,.}
\end{array}
\ee
In particular, the results in Fig.\,\ref{fig:I3Ising_both} correspond to $\alpha=0.25$ (left panel) and $\alpha=2$ (right panel), where the dots denote the values of $I^{(n)}_{A_1,A_2,A_3}$ for $n\in \{2,3,4\}$.
Unfortunately, with the computational resources at our disposal, we could not  compute R\'enyi entropies for higher values of $n$.
Indeed, besides the problem of computing the Riemann theta function numerically for large period matrices, the additional obstacle occurring for the Ising model is that the number of elements in the sum (\ref{FIMGeneralN}) grows exponentially with $n$.
Given the few $n$'s available, only few rational interpolations can be employed to approximate the analytic continuation to $n=1$ and they are depicted in Fig.\,\ref{fig:I3Ising_both} through solid and dashed lines (in general we never use $(p,q)=(0,1)$ because is often not well-behaved).
It is interesting to observe that the three different rational interpolations provide the same extrapolation to $n=1$ for a large range of $x_2$ (they differ when $x_2$ is close to 1). 
Since, to our knowledge, numerical results about $I_{A_1,A_2,A_3}$ for the Ising model are not available in the literature, the curves in  Fig.\,\ref{fig:I3Ising_both} are predictions that would be interesting to test through other methods. 
\\
In order to check the reliability of the numerical method, we have performed rational interpolations considering only $n\in \{3,4\}$ to extrapolate the value at $n=2$, which is known analytically. Since only two points are available, only the rational interpolation with $(p,q)=(1,0)$ can be done, which is given by the dot-dashed curve in Fig.\,\ref{fig:I3Ising_both}. Despite the roughness of the extrapolation due to the few input points, the agreement with the expected values computed with the analytic expression (black dots) is very good.


\section{Entanglement negativity of two disjoint intervals}
\label{sec neg 2int}

In this section we consider the logarithmic negativity of two disjoint intervals for the non compact massless free boson, whose analytic formula is not known.

The method to compute the logarithmic negativity $\mathcal{E}$ in quantum field theory and in conformal field theory has been described in \cite{cct-neg-letter, cct-neg-long} (see \cite{cct-neg-T} for the finite temperature case) and we refer to these papers for all the details and the discussion of further cases.
In order to briefly mention the main idea, let us consider a subsystem $A=\cup_{i=1}^N A_i$ made by $N$ disjoint intervals $A_i =[u_i, v_i]$.
The traces $\Tr \rho_A^n$ in CFT are given by the correlators of twist fields in (\ref{Tr rhoA Nint cft}). 
 Denoting by $A_0 \varsubsetneq A$ a set of $N_0 < N$ disjoint intervals among the ones in $A$ and by $\rho_A^{T_0}$ the partial transpose of $\rho_A$ with respect to $A_0$, we have that $\Tr (\rho_A^{T_0} )^{n}$ in CFT is the correlation function of twist fields obtained by placing $\mathcal{T}_n$ in $u_i$ and $\bar{\mathcal{T}}_n$ in $v_i$ when $A_i \in A\setminus A_0$, and 
$\bar{\mathcal{T}}_n$ in $u_i$ and $\mathcal{T}_n$ in $v_i$ when $A_i \in A_0$.
The corresponding logarithmic negativity $\mathcal{E}$, which measures the entanglement between $A_0$ and $A\setminus A_0$, can be computed by considering the sequence of the even integers $n_e$ and taking the replica limit (\ref{replica limit neg}).
Configurations containing adjacent intervals are obtained as limiting cases and the fields $\mathcal{T}_n^2$ and $\bar{\mathcal{T}}_n^2$ occur.

In the simplest example, starting from two disjoint intervals $A=A_1 \cup A_2$, whose endpoints are ordered as $u_1 < v_1< u_2<v_2$ like in \S\ref{sec MI}, one considers e.g. the partial transpose with respect to $A_2$.
In this case we have that \cite{cct-neg-letter, cct-neg-long}
\bea
\label{TrRhoNeg}
{{\rm Tr}}( \rho_A^{T_2})^n 
&=&
\langle  \mathcal{T}_n(u_1)  \bar{\mathcal{T}}_n(v_1)
\bar{\mathcal{T}}_n(u_2)  \mathcal{T}_n(v_2)  \rangle 
\\
\label{G function def}
\rule{0pt}{.8cm}
&=&
c_n^2
\left[\frac{(u_2-u_1)(v_2-v_1)}{(v_1-u_1)(v_2-u_2)(u_2-v_1)(v_2-u_1)}\right]^{2\Delta_n} 
\mathcal{G}_{n}(x)\,,
\eea
where $x \in (0,1)$ is the four-point ratio (\ref{x ratio def}) and $\Delta_n$ has been introduced in (\ref{twist 2point func}).
Since (\ref{TrRhoNeg}) is obtained from (\ref{renyi cft 2int}) by exchanging $\mathcal{T}_n \leftrightarrow \bar{\mathcal{T}}_n$ for the endpoints of $A_2$, the function $\mathcal{G}_{n}$ in (\ref{G function def}) is related to the function $\mathcal{F}_{2,n}$ in (\ref{F2n def}) as follows
\be
\label{Gn through Fn}
\mathcal{G}_{n}(x)
\equiv
(1-x)^{4\Delta_n} \, \mathcal{F}_{2,n}\big(x/(x-1)\big)\,,
\ee
where we remark that $x/(x-1) \in (-\infty, 0)$.
Plugging (\ref{Gn through Fn}) into (\ref{G function def}) and taking the replica limit (\ref{replica limit neg}) of the resulting expression, since $\Delta_1 =0$ and $c_1=1$, we find that the logarithmic negativity of two disjoint intervals in CFT is given by
\be
\label{neg limit G_n}
\mathcal{E}(x)
= \lim_{n_e\rightarrow1}  \log \mathcal{G}_{n_e}(x)
=\lim_{n_e\rightarrow1}
\log \mathcal{F}_{2,n_e}\big(x/(x-1)\big)\,,
\ee
telling us that the logarithmic negativity is scale invariant, being a function of the ratio $x$ only.
In order to get rid of the prefactor in (\ref{G function def}), it is convenient to consider the following ratio
\be
\label{RatioNeg}
\widetilde{R}_n
\equiv \frac{{{\rm Tr}}( \rho_A^{T_2})^n}{{{\rm Tr}}( \rho_A)^n}
=
\frac{\mathcal{G}_{n}(x)}{\mathcal{F}_{2,n}(x)}
= 
(1-x)^{4\Delta_n}\,
\frac{\mathcal{F}_{2,n}\big(x/(x-1)\big)}{\mathcal{F}_{2,n}(x)}\,,
\ee
where (\ref{Gn through Fn}) has been employed in the last step. 
Since $\mathcal{F}_{2,1}(x)=1$ for $x\in (0,1)$ because of the normalization of $\rho_A$, the logarithmic negativity can be found also by taking the replica limit of (\ref{RatioNeg}), namely
\be
\label{tilde Rn def}
\mathcal{E}(x) =\, \log \lim_{n_e\rightarrow1} \widetilde{R}_{n_e}(x)\,,
\ee
Notice that, since for $n=2$ we have that $\mathcal{T}_2 = \bar{\mathcal{T}}_2$, one concludes that $\widetilde{R}_2 =1$ identically.

The simplest model we can deal with for which analytic expressions for ${{\rm Tr}}( \rho_A^{T_2})^n$ are available in the literature is the non compact free massless boson.
For this model it has been found that \cite{cct-neg-letter, cct-neg-long}
\be
\label{tildeR def}
\widetilde{R}_{n}(x)
=
(1-x)^{(n-1/n)/3}
\left[\frac{\prod_{k=1}^{n-1}F_{k/n}(x)F_{k/n}(1-x)}{
\prod_{k=1}^{n-1}
\textrm{Re}\big(F_{k/n}(\frac{x}{x-1})\bar{F}_{k/n}(\frac{1}{1-x})\big)}\right]^{1/2}.
\ee
When $n=n_e$ is even, it could be convenient to isolate the term $k/n=1/2$ in the product in order to get rid of the square root in the remaining part of the product because of the symmetry $k \leftrightarrow n-k$ in $F_{k/n}$.
Notice that when $n=2$ we have that $\widetilde{R}_{2}(x)=1$ identically.  

\begin{figure}[t]
\vspace{.3cm}
\hspace{-1cm}
\includegraphics[width=17.5cm]{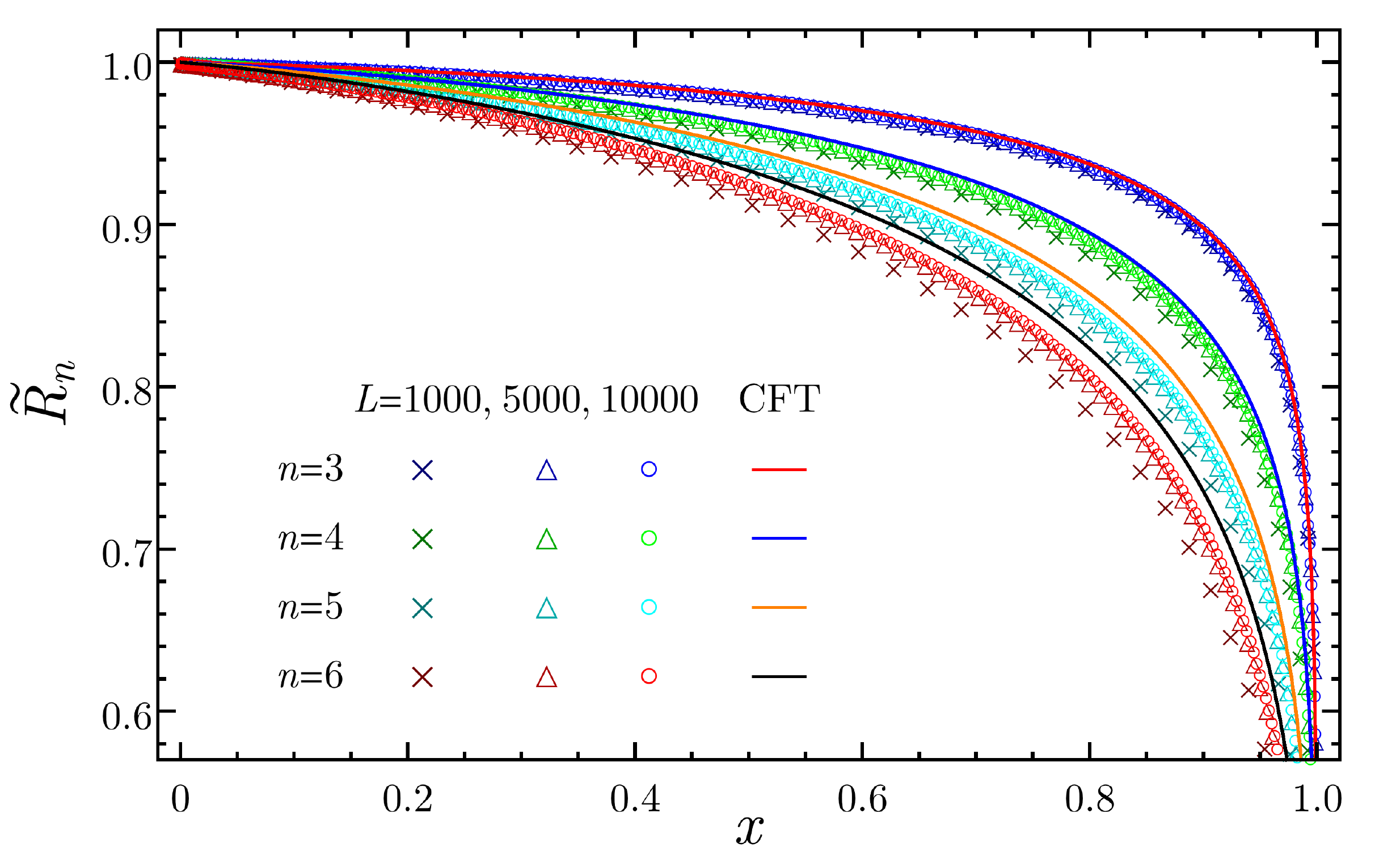}
\vspace{-.7cm}
\caption{
The ratio $\widetilde{R}_n(x)$ in (\ref{RatioNeg}) for the non compact boson. The data points come from the periodic harmonic chain with $\omega L =10^{-5}$, while the curves are given by CFT formula (\ref{tildeR def}).
}
\label{fig:Rn_neg}
\end{figure}

\begin{figure}[t]
\vspace{.3cm}
\hspace{-1cm}
\includegraphics[width=17.5cm]{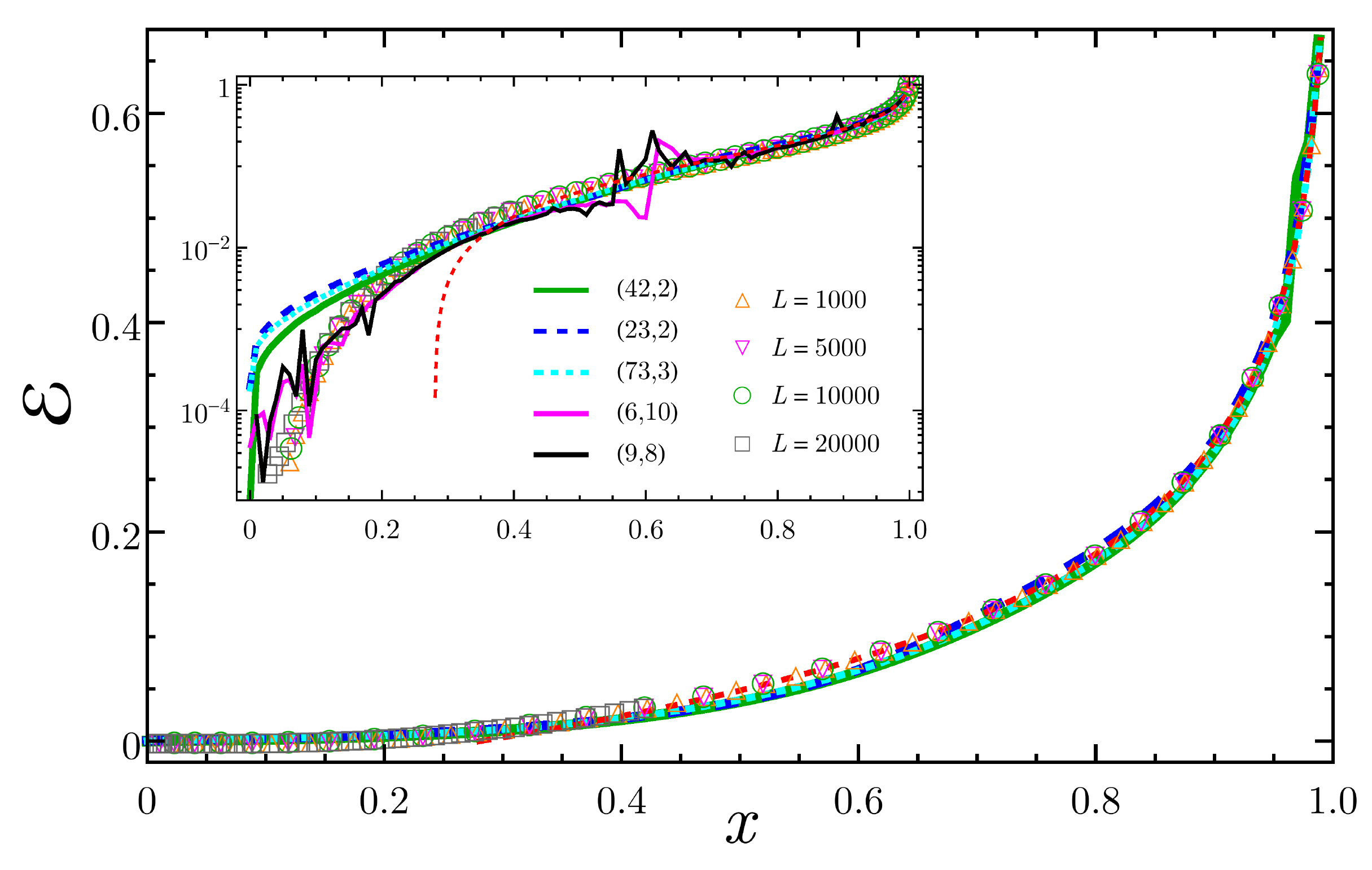}
\vspace{-.7cm}
\caption{
Logarithmic negativity of two disjoint intervals for the non compact boson (\ref{eq:LN})  as function of the four-point ratio $x$.
The dots are numerical data obtained for the periodic harmonic chains with $\omega L = 10^{-5}$ and increasing total lengths.
All data collapse on the same curve, which corresponds to the continuum limit.
The red dashed curve is the analytic continuation found in \cite{cct-neg-long} in the regime $x \to 1^{-}$.
The remaining curves are extrapolations obtained from different rational interpolations having $(p,q)$ indicated.
In the inset we show the same plot in logarithmic scale in order to highlight the behaviour of the different extrapolated curves when $x\sim 0$.
}
\label{fig:Rl_NegAll}
\end{figure}

In Fig.\,\ref{fig:Rn_neg} we compare the CFT result (\ref{tildeR def}) for $\widetilde{R}_{n}(x)$ with the corresponding quantity computed for the periodic harmonic chain (\ref{HC ham}), where ${{\rm Tr}}( \rho_A^{T_2})^n$ is computed through the correlators $\langle q_r q_s \rangle$ and $\langle p_r p_s \rangle$ as explained in \cite{entHC}.
Notice that we have improved this check with respect to \cite{cct-neg-long}, indeed 
the data in Fig.\,\ref{fig:Rn_neg} correspond to chains whose total length $L$ is significantly larger than the ones considered in \cite{cct-neg-long}, where $L \leqslant 300$. 
All the data reported in the figure have $\omega L =10^{-5}$. We have considered also harmonic chains with $\omega L =10^{-3}$ and $L=10000$, finding the same results reported in Fig.\,\ref{fig:Rn_neg}  for $L=10000$.
For $n=3$ the agreement is very good, while it gets worse as $n$ increases.
This is expected because of the unusual corrections to the scaling \cite{unusual-corr}.

It is more convenient to consider (\ref{neg limit G_n}) than (\ref{tilde Rn def}) for the computation of the replica limit, and for the logarithmic negativity of the non compact boson we have that \cite{cct-neg-long}
\be
\label{eq:LN}
\mathcal{E}(x)
\,=\, 
- \frac{1}{2} \log\big[K(x) K(1-x)\big] 
- \frac{3}{8} \log (1-x) 
+\log(\pi/2)
-
\lim_{n_e\rightarrow1}   \sum_{k=1}^{n_e/2-1} 
\log G_{k/n_e}(x)\,,
\ee
where
\be
\label{eq:LN_G}
G_\beta(x) \,\equiv \,
_2F_1(\beta,\beta,1;x) 
\left[
\,\frac{\Gamma(1-2\beta)}{\Gamma(1-\beta)^2}
\,(1-x)^\beta\, _2F_1(\beta,\beta,2\beta;1-x) 
-  (\beta \leftrightarrow 1-\beta) \right] ,
\ee
being $K(x)$ the elliptic integral of the first kind. 
The sum in (\ref{eq:LN}) is defined for $n_e \geqslant 4$ and for $n_e=2$ that term is zero. 
The analytic continuation in (\ref{eq:LN}) is not known for the entire range $x\in (0,1)$.
In \cite{cct-neg-long} the analytic continuation has been found for the regime $x \to 1^{-}$, obtaining an expression that surprisingly works down to $x\sim 0.3$ (see the dashed red curve in Fig.\,\ref{fig:Rl_NegAll}).

Here we numerically extrapolate $\mathcal{E}(x)$ through the formula (\ref{eq:LN}) by using  the rational interpolation method, which has been discussed in \S\ref{RIsection} and employed in the previous sections for the entanglement entropy of disjoint intervals.
It is worth remarking that, since the replica limit (\ref{replica limit neg}) for $\mathcal{E}(x)$ involves only even $n$'s, to perform a rational interpolation characterized by some $(p,q)$ we need higher values of $n$ with respect to the ones employed for the entanglement entropy in the previous sections. 
In particular, for the logarithmic negativity $p+q+1 \leqslant n_{e, \textrm{\tiny max}}/2$.

In Fig.\,\ref{fig:Rl_NegAll} we report the extrapolations found for some values of $(p,q)$.
Since the numerical data from the harmonic chain are accurate enough to provide the curve in the continuum limit that should be found through the analytic continuation (\ref{eq:LN}), we can check the reliability of our numerical extrapolations against them.
For the non compact boson the expression (\ref{eq:LN_G}) is not difficult to evaluate numerically. Thus, we can deal with high values of $n$ and therefore we have many possibilities for $(p,q)$.
It turns out that an accurate extrapolation for the logarithmic negativity requires high values of $p$ and $q$, in particular for the regime of small intervals $x\sim 0$ (see Fig.\,\ref{fig:Neg2} in \S\ref{RIsection} for extrapolations having low $p$ and $q$).
As already remarked in \cite{cct-neg-letter, cct-neg-long}, the behaviour of $\mathcal{E}(x)$ when $x\sim 0$ is not power-like.
We observed, as a general behaviour, that increasing $q$ leads to extrapolations which are closer to the numerical data, but spurious fluctuations or even singularities in some regimes of $x$ can occur
(see the black and magenta curves in the inset of Fig.\,\ref{fig:Rl_NegAll}, and the dashed magenta and cyan curves in Fig.\,\ref{fig:Neg2}).
This happens whenever one of the $q$ poles of the rational function is close to the range $(1, n_{\textrm{\tiny max}})$ of the interpolated data and not too far from $n=1$ (it may be real or have a small imaginary part).
More details are reported in \S\ref{RIsection}.
Taking low $q$'s, one usually gets smooth curves but even high values of $p$'s are not sufficient to capture the behaviour of $\mathcal{E}(x)$ when $x\sim 0$.

Thus, the logarithmic negativity is more difficult to find through the rational interpolation method than the entanglement entropy.
Indeed, while for the latter one few R\'enyi entropies are enough to capture the expected result in a stable way, for the logarithmic negativity more input data are needed to reproduce the regime of distant intervals. 
Maybe other numerical methods are more efficient.
It is worth remarking that the fact that high values of $n$'s in ${{\rm Tr}}( \rho_A^{T_2})^n $ are required to perform accurate extrapolations of the logarithmic negativity leads to a computational obstacle whenever $\mathcal{G}_{n}(x)$ in (\ref{G function def}) is written in terms of Riemann theta functions, like for the compact boson \cite{cct-neg-long}  and for the Ising model \cite{a-13, ctt-13}.
Given our computational resources, we have not been able to deal with those analytic expressions for $n$ high enough to guarantee convincing extrapolations.

\section{Conclusions}

The analytic continuations leading to analytic expressions for the entanglement entropy and the logarithmic negativity of disjoint regions can be very difficult to perform, even for simple CFTs. 
In this paper we studied this problem numerically for the CFTs given by the free massless boson (compactified or in the decompactification regime) or by the Ising model, where $ \Tr\rho_A^n$ for a generic number of disjoint intervals \cite{cct-09, cct-11, ctt-14} and $\Tr (\rho_{A_1\cup A_2}^{T_2} )^n$ are known analytically \cite{cct-neg-letter, cct-neg-long, a-13, ctt-13}.

The numerical extrapolations have been performed through a method based on rational interpolations, which has been first employed in this context by \cite{ahjk-14}.
Its reliability has been checked by reproducing the existing results  found from the corresponding lattice models through various techniques like exact diagonalizations \cite{fps-08, ctt-14} and Tree Tensor Networks \cite{atc-10}.
In our analysis, we observed that for the entanglement entropy one finds the same curve through different extrapolations already with small values of the degrees $p$ and $q$ of the polynomials occurring in the numerator and in the denominator respectively of the rational interpolation. Instead, for the logarithmic negativity higher values of $p$ and $q$ are needed for the regime of distant intervals, where it falls off faster than any power. 
Extrapolations having higher values of $q$ are more efficient in providing the expected result, but they can show some spurious behaviour in some parts of the domain.
Our numerical analysis has been limited both by our computational resources (in the evaluation of the Riemann theta functions for large matrices) and by the features of the model (e.g. for the logarithmic negativity of distant intervals).
These obstacles prevented us to treat some interesting cases like the logarithmic negativity of two disjoint intervals for the compact boson and for the Ising model because high values of $n$ are needed to get convincing extrapolations. 
We remark that lattice results for $\mathcal{E}(x)$ have been found in  \cite{ctt-13} for the Ising model through Tree Tensor Networks, while for the compact boson they are not available in the literature (see \cite{AlbaLauchlin-neg} for $\widetilde{R}_3$ obtained through Quantum Monte Carlo).

When singularities in $n$ occur (see e.g. \cite{moore-2d}), the numerical method adopted here is expected to fail.
As for the one dimensional systems that have been considered, given the good agreement with the lattice results, a posteriori  we expect that there are no singularities in the ranges of $n$ that have been explored.

The rational interpolation method has been also employed to address some cases whose corresponding lattice results are not available in the literature (e.g. the $U(1)$ gauge theory in $2+1$ dimensions has been studied in \cite{ahjk-14} and the case of three disjoint intervals for the Ising model in \S\ref{sec tripartite}).
Thus, it is a useful tool that could be used in future studies to find numerically the entanglement entropy and the logarithmic negativity of disjoint regions (or for single regions whenever the analytic continuation is difficult to obtain) for other interesting situations like e.g. for CFTs in higher dimensions \cite{cft-high-dims} and in the context of the holographic correspondence  \cite{RT,headrick,hol}.


\section*{Acknowledgments}

We are grateful to Pasquale Calabrese and Luca Tagliacozzo for useful comments on the draft. 
We thank all the authors of \cite{fps-08, atc-10} for allowing us to use their numerical results.
ET has been supported by the ERC under  Starting Grant  279391 EDEQS.

\begin{appendices}

\section*{Appendices}

\section{Rational interpolations}
\label{RIsection}

In this appendix we discuss the numerical method that we have employed throughout the paper, which is based on rational interpolations, and the issues we encountered to address the replica limits for the entanglement entropy and negativity considered in the main text. 
Its use in this context has been first suggested in \cite{ahjk-14}.

The rational interpolation method consists in constructing a rational function which interpolates a finite set of given points labeled by a discrete variable. 
Once the rational function is written, one simply lets the discrete variable assume all real values. The needed extrapolation is found by just evaluating the rational function obtained in this way for the proper value of the variable.  

For the quantities we are interested in, the discrete variable is an integer number $n$.
As a working example, let us consider the case of two disjoint intervals, where the variable $x\in (0,1)$ characterizes the configuration of intervals. 
For any integer $n\geqslant 2$ we have a real function of $x$ and typically we have access only to $n \leqslant n_{\textrm{\tiny max}}$ for computational difficulties.
The rational function interpolating the given data reads
\be
\label{RIDef}
W^{(n)}_{(p,q)}(x) 
\equiv 
\frac{P(x;n)}{Q(x;n)}
\equiv 
\frac{a_0(x)+ a_1(x) n+a_2(x)n^2+\cdots +a_p(x)n^p}{b_0(x)+b_1(x)n+b_2(x)n^2+\cdots +b_q(x)n^q} \,,
\ee
being $p\equiv \textrm{deg}(P)$ and $q\equiv \textrm{deg}(Q)$ the degrees of the numerator and of the denominator respectively as polynomials in $n$.
The extrapolations are performed pointwise in the domain $x\in (0,1)$.
Thus, for any given $x\in (0,1)$, in (\ref{RIDef}) we have $p+q+2$ coefficients to determine. 
Nevertheless, since we can divide both numerator and denominator by the same number fixing one of them to 1, the number of independent parameters to find is $p+q+1$.
Once the coefficients in (\ref{RIDef}) have been found, the extrapolation is easily done by considering $n$ real and setting it to the needed value. 
It is important to stress that, having access only to a limited number $m$ of data points, we can only perform rational interpolations whose degrees $(p,q)$ are such that $p+q+1 \leqslant m$.
This method is implemented in \emph{Wolfram Mathematica} through the \emph{Function Approximations} package and the command \emph{RationalInterpolation}.

\begin{figure}[t]
\vspace{.3cm}
\hspace{-.0cm}
\includegraphics[width=15cm]{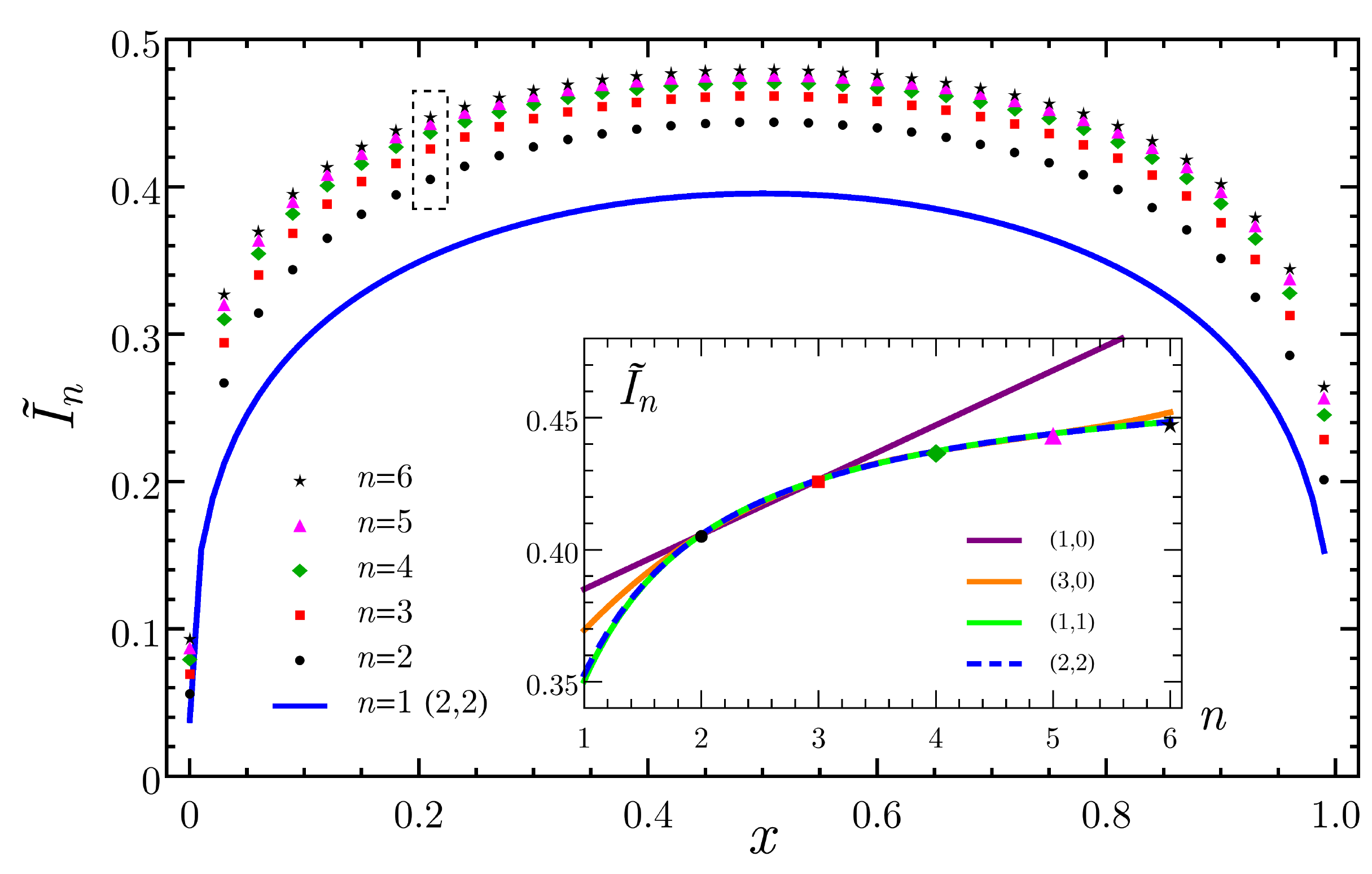}
\vspace{-.1cm}
\caption{
The quantity $\tilde{I}_n$ in (\ref{RenyiMI cft}) and the corresponding $n \to 1$ limit (\ref{replica limit MI cft}) for the compact boson ($c=1$) with $\eta=0.295$.
The blue line is the extrapolation $n=1$ of the rational interpolation with $(p,q)=(2,2)$ obtained through the analytic expressions given by (\ref{F 2int compact}) and (\ref{tau2 def}) with $2 \leqslant n \leqslant 6$, whose values for $\tilde{I}_n$ are shown by points for some values of the four point ratio $x$.
In the inset, considering the configuration having  $x=0.2101$ (highlighted by the dashed rectangle in the main plot), we show $\tilde{I}_n$ as function of $n$ for rational interpolations having different $(p,q)$. The extrapolations having $q>0$ capture the expected value better than the ones having $q=0$.
}
\label{fig:RI_Ex_Inset}
\end{figure}

\begin{figure}[t]
\vspace{.3cm}
\hspace{-.0cm}
\includegraphics[width=15cm]{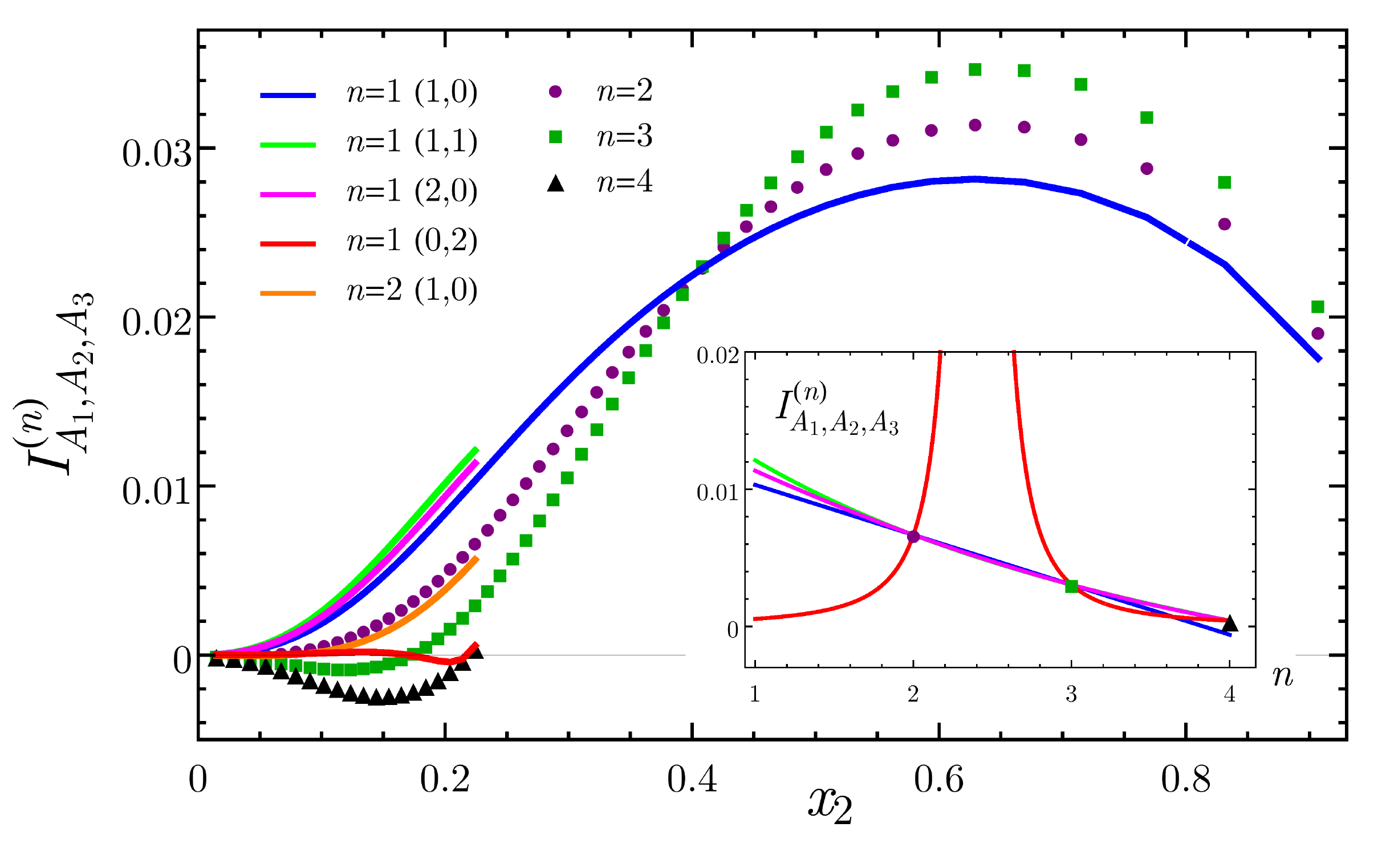}
\vspace{-.3cm}
\caption{
Three disjoint intervals: The quantity $I^{(n)}_{A_1, A_2, A_3}$ in (\ref{I3n def}) for the compact boson, computed through (\ref{RNn cft N=3}) and (\ref{F Nint compact boson}) for $n \geqslant 2$. 
Our limited computational power in evaluating Riemann theta functions for large matrices prevented us to consider $n=4$ in the whole range of configurations and this limits also the possible rational interpolations that can be employed. 
The blu line is the extrapolation found by using only $n\in \{2,3\}$, which should not be considered as a prediction because more $n$'s are needed to find stable extrapolations. 
The orange line is a check of the method for $n=2$: the fact that the expected points are not precisely recovered is due to low number of $n$'s ($n\in \{3,4\}$) available. 
In the inset, considering the configuration having  $x_2=0.224$, we show $I^{(n)}_{A_1, A_2, A_3}$ as function of $n$ for rational interpolations having different $(p,q)$. The rational interpolation with $(p,q) = (0,2)$ (red line) shows a bad behaviour and the extrapolation to $n=1$ cannot be trusted; indeed, the red curve in the main plot is different from the other extrapolations.
}
\label{fig:RI_Ex2_Inset}
\end{figure}

In Fig.\,\ref{fig:RI_Ex_Inset} we consider an explicit example where we extrapolate the $\tilde{I}_1(x)$ in (\ref{replica limit MI cft}) of the compact boson ($c=1$) for a particular value of the compactification radius corresponding to $\eta=0.295$ (see also Fig\,\ref{fig:RI_FPS}).
For $n\geqslant 2$ the analytic expressions are (\ref{RenyiMI cft}) and (\ref{F 2int compact}) and we take into account $2\leqslant n \leqslant 6$ only (in Fig.\,\ref{fig:RI_FPS} we employ also $n=7$).
Given these data, we can perform rational interpolations with $p+q+1\leqslant 5$.
The blue curve in Fig.\,\ref{fig:RI_Ex_Inset} is the extrapolation to $n=1$ of the rational interpolation with $(p,q)=(2,2)$. 
We find it instructive to describe the details for a specific value of $x$.
Let us consider, for instance, a configuration corresponding to $x = \tilde{x} \equiv 0.2101$ (see the dashed rectangle in Fig.\,\ref{fig:RI_Ex_Inset}).
First one has to compute the rational interpolation with $(p,q)=(2,2)$, then the limit $n \to 1$ must be taken. For these two steps, we find respectively
\be
W^{n}_{(2,2)}(\tilde{x})=\frac{0.358 - 0.480\,n + 
 3.689\,n^2}{1 + 1.347\,n + 7.870\,n^2}\,, 
 \qquad 
 \lim_{n\rightarrow1}W^{n}_{(2,2)}(\tilde{x})=0.349\,.
\ee
In the inset of Fig.\,\ref{fig:RI_Ex_Inset} we show how adding more data improves the final extrapolation and how it becomes stable. Focusing again on $x = \tilde{x}$, we can start by taking only $n \in \{2,3\}$, which allow  to perform a rational interpolation with $(p,q)=(1,0)$ (a line). 
Since rational interpolations having $p=0$ often provide wrong predictions, we prefer to avoid them, if possible. 
The extrapolation to $n=1$  corresponding to $(p,q)=(1,0)$ cannot be trusted and therefore we consider four input data $n \in \{2,3,4,5\}$ which allow to consider a rational interpolation with, for instance, $(p,q)=(3,0)$ and also $(p,q)=(1,1)$. These two different rational interpolations do not provide the same extrapolation to $n=1$ and therefore we must take into account more input data.
Considering $2\leqslant n \leqslant 6$ we can choose also $(p,q)=(2,2)$ finding that the corresponding rational interpolation basically coincides with the one with $(p,q)=(1,1)$ (their difference is of order $10^{-3}$).
Thus, the extrapolation to $n=1$ obtained with $(p,q)=(2,2)$ is quite stable.
Repeating this analysis for the whole range of $x\in (0,1)$, one can find the blue curve in Fig.\,\ref{fig:RI_Ex_Inset}.
As a further check, in Fig.\,\ref{fig:RI_FPS} we have used $(p,q)=(3,2)$ using more input data, finding that the final extrapolation is basically the same.
Plots like the one shown in the inset of Fig.\,\ref{fig:RI_Ex_Inset} are very useful to understand the stability of the extrapolation to $n=1$.
Increasing the values of $p$ and $q$ in the rational interpolations leads to more precise extrapolations, as expected.
Rational interpolations with $q>0$ provide extrapolations which are closer to the expected value with respect to the ones with $q=0$. 
When $q$ is strictly positive, $q$ poles occur in the complex plane parameterized by $n \in \mathbb{C}$. 
Nevertheless, if these poles are far enough from the real interval $(1, n_{\textrm{\tiny max}})$ containing all the $n$'s employed as input data for the interpolation, the extrapolations to $n=1$ are reliable.
Increasing $q$, we have higher probability that one of the poles is close to the region of interpolation, spoiling the extrapolation.
Plotting $W^n_{(p,q)}(x)$ as function of $n$ is useful to realize whether this situation occurs (see the inset of Fig.\,\ref{fig:RI_Ex2_Inset} for an explicit example).

\begin{figure}[t]
\vspace{.3cm}
\hspace{-.0cm}
\includegraphics[width=15cm]{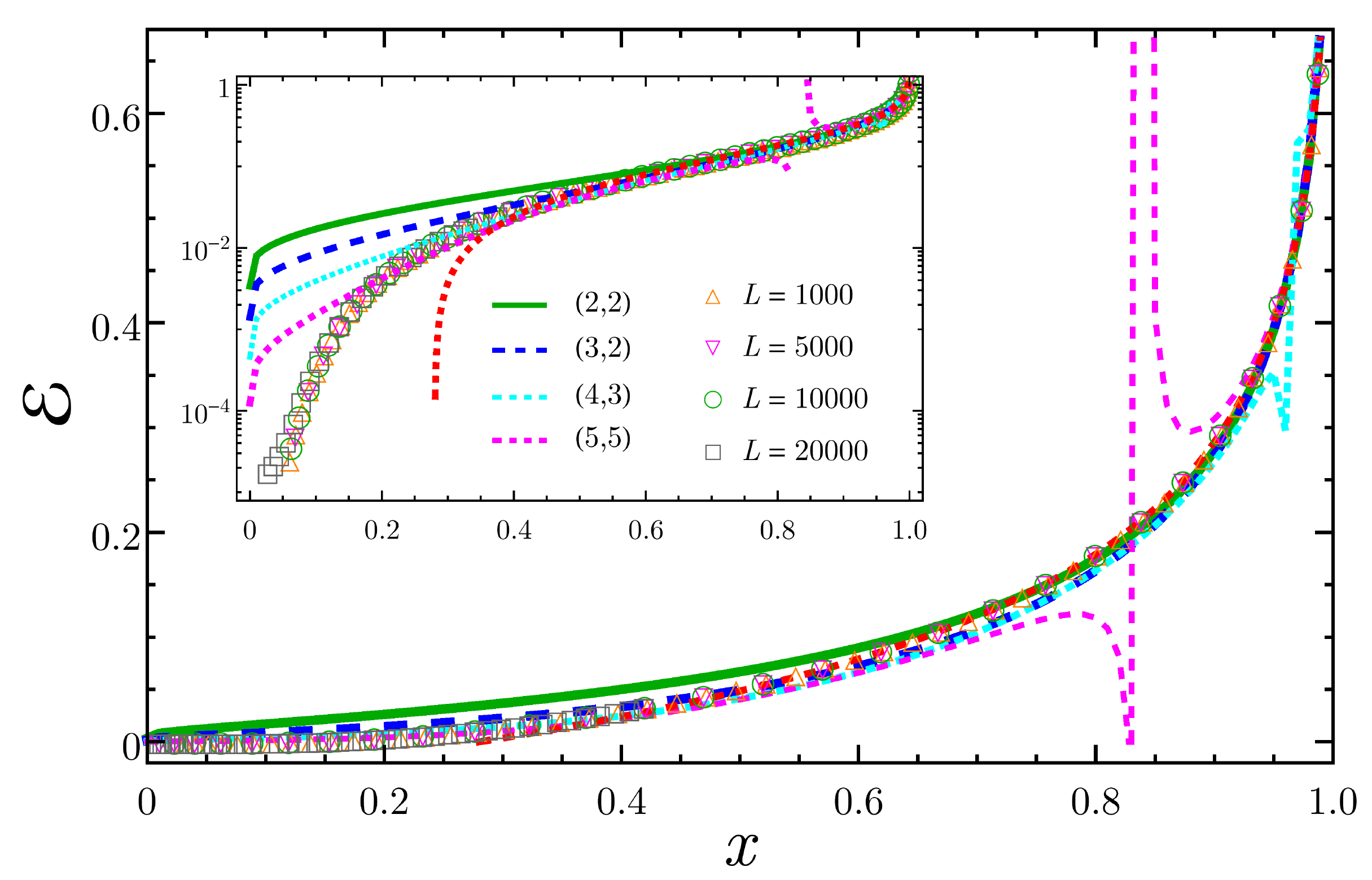}
\vspace{-.1cm}
\caption{
Logarithmic negativity of two disjoint intervals for the non compact boson: Extrapolations having low values of $p$ and $q$. This plot should be compared with Fig.\,\ref{fig:Rl_NegAll}, where higher values of  $p$ and $q$ have been considered.
Increasing $q$ improves the extrapolation but in some regimes of $x$ wrong results can be found.
The dashed red curve is the analytic continuation for the regime $x \to 1^{-}$ found in \cite{cct-neg-long}, while the points are obtained through a periodic harmonic chain (\ref{HC ham}) with $L$ sites.
}
\label{fig:Neg2}
\end{figure}

The issue of evaluating Riemann theta functions which involve large matrices becomes important when we want to compute $I_{A_1,A_2,A_3}$ (see (\ref{I3 def}) and (\ref{I3n def})) for a compact boson. Indeed, $\mathcal{F}_{3,n}(\boldsymbol{x})$ in (\ref{RNn cft N=3}) is given by (\ref{F Nint compact boson}) for $N=3$ and therefore the matrix occurring in the Riemann theta function is $2g \times 2g$ with $g=2(n-1)$.
Given our computational power, we computed $I^{(n)}_{A_1,A_2,A_3}$ for $n\in \{2,3\}$ for all the needed configurations of intervals, while for $n=4$ we got results only for small intervals. 
In Fig.\,\ref{fig:RI_Ex2_Inset} we show our data and some numerical extrapolations. 
In the whole range of $x_2$ we performed only the rational interpolation with $(p,q)=(1,0)$ (blue line) because only two input data are available, while for $x_2 \in (0, 0.22)$, where also $n=4$ is available, we could employ higher values of $p$ and $q$.
When we have more extrapolations, unfortunately they do not overlap, indicating that we cannot trust these curves to give a prediction, even if they are quite close. Another indication that $n=4$ is not enough to get a precise extrapolation comes from the fact that, given the data with $n \in \{3,4\}$ and extrapolating to $n=2$ (orange curve in Fig.\,\ref{fig:RI_Ex2_Inset}) we did not recover exactly the expected values (purple circles) found with the analytic expressions.
In the inset we focus on a configuration of three intervals corresponding to $x_2=0.224$ and show the dependence of $I^{(n)}_{A_1,A_2,A_3}$ on $n$ for various $(p,q)$. 
While the extrapolations to $n=1$ associated to $(1,0)$ (for this one only $n\in \{2,3\}$ have been used), $(1,1)$ and $(2,0)$ are very close, the one corresponding to $(p,q)=(0,2)$ provides a completely different extrapolation to $n=1$.
Considering the two poles of the interpolating function in the regime of $x_2$ where also $n=4$ is available, 
we find that they are real and at least one of them is inside the domain $n\in(1,4)$.
Thus, the function cannot be considered a good approximation of the true analytic continuation and the extrapolation cannot be trusted.
This behaviour does not occur for the case considered in the inset of Fig.\,\ref{fig:RI_Ex_Inset}. Thus, it is useful to plot the $n$ dependence of the functions obtained through the rational interpolation method in order to check the occurrence of singularities that could lead to wrong extrapolations.

We find it instructive to discuss some details about the extrapolations of the logarithmic negativity of two disjoint intervals (see \S\ref{sec neg 2int}).
The simplest case we can deal with is the non compact boson and the replica limit to perform for this model is (\ref{eq:LN}). 
The analytic expression (\ref{eq:LN_G}) contains only hypergeometric functions and therefore it can be evaluated for high values of $n$.
Some extrapolations performed through the rational interpolation method explained above are shown in Figs.\,\ref{fig:Rl_NegAll} and \ref{fig:Neg2}.
The first difference between the logarithmic negativity and the mutual information in the extrapolation process is that for the former quantity we need to consider higher values of $p$ and $q$ with respect to the latter one to recover the expected result.
Moreover, in the regime of small intervals or large separation (i.e. $x \sim 0$), where  the logarithmic negativity  falls off to zero faster than any power, it is very difficult to capture its behaviour in a clean way, despite the high values of $p$ and $q$. 
In Fig.\,\ref{fig:Neg2} we show some extrapolations characterized by low values of $p$ and $q$. The most difficult regime to capture is the one with $x\sim 0$. Thus, in Fig.\,\ref{fig:Rl_NegAll} we show some extrapolations having higher values of $p$ and $q$.
Comparing the curves in these figures, one observes that with low $q$'s it is difficult to capture the regime of small $x$, even for very high values of $p$. 
Increasing $q$, the agreement slightly improves for small $x$, but, as already remarked, it is more probable that the singularities of the rational interpolation fall close to the domain of the interpolated data.
For example, in the case of the dashed magenta curve of Fig.\ref{fig:Neg2}, all the poles of the rational function are real. Varying the parameter $x$, they move on the real axis and, whenever one of them comes close to the interpolation region $(1, n_{\textrm{\tiny max}})$ and it is not too far from $n=1$, the extrapolated function to $n=1$ cannot be trusted as approximation of the true analytic continuation.
This leads to fluctuations or singularities in the extrapolation curve as function of $x$ (e.g. see also the dashed cyan curve in Fig.\,\ref{fig:Neg2} and the black and magenta curves in Fig.\,\ref{fig:Rl_NegAll}).

\end{appendices}


\end{document}